\documentclass[epj]{svjour}
\usepackage{graphicx,color}
\usepackage[justification=justified,tableposition=top,labelfont=bf,margin=1.5cm]{caption} 
\usepackage{subcaption}
\usepackage{amssymb}
\usepackage{amsmath,slashed}
\usepackage[bookmarks,pdfhighlight=/O,pdfstartview=FitH]{hyperref}

\newcommand{\erw}[1]{\ensuremath { %
    \left \langle {#1} \right \rangle}}
\newcommand{\MeV}{\ensuremath{\mathrm{MeV}}}

\newcommand{\fm}{\ensuremath{\mathrm{fm}}}
\newcommand{\ii}{\ensuremath{\mathrm{i}}}
\newcommand{\fslash}{\slashed}
\newcommand{\dd}{\ensuremath{\mathrm{d}}}
\newcommand{\bvec}[1]{\ensuremath{\boldsymbol{#1}}}

\newcommand{\vv}[2]{\begin{pmatrix} #1\\#2 \end{pmatrix}}
\newcommand{\R}{\ensuremath{\mathbb{R}}}

\begin{document}

\title{Kinetics of the chiral phase transition in a linear $\sigma$ model}

\author{Christian Wesp\inst{1}\thanks{\email{cwesp@th.physik.uni-frankfurt.de}} \and
  Hendrik van
  Hees\inst{1,2}\thanks{\email{hees@th.physik.uni-frankfurt.de}} \and
  Alex
  Meistrenko\inst{1}\thanks{\email{meistrenko@th.physik.uni-frankfurt.de}}
  \and
  Carsten Greiner\inst{1}\thanks{\email{carsten.greiner@th.physik.uni-frankfurt.de}}}

\institute{Institut f{\"u}r theoretische Physik, Goethe-Universit{\"a}t
  Frankfurt am Main, Max-von-Laue-Stra{\ss}e 1, 60438 Frankfurt, Germany
  \and
  Frankfurt Institute for Advanced Studies, Ruth-Moufang-Stra{\ss}e 1, 60438 Frankfurt, Germany}

\abstract{We study the dynamics of the chiral phase transition in a
  linear quark-meson $\sigma$ model using a novel approach based on
  semiclassical wave-particle duality. The quarks are treated as test
  particles in a Monte-Carlo simulation of elastic collisions and the
  coupling to the $\sigma$ meson, which is treated as a classical field,
  via a kinetic approach motivated by wave-particle duality. The
  exchange of energy and momentum between particles and fields is
  described in terms of appropriate Gaussian wave packets. It has been
  demonstrated that energy-momentum conservation and the principle of
  detailed balance are fulfilled, and that the dynamics leads to the
  correct equilibrium limit. First schematic studies of the dynamics of
  matter produced in heavy-ion collisions are presented.}
\maketitle
\section{Introduction}
\label{sec:1}

The exploration of the phase diagram of strongly interacting
matter \cite{Friman:2011zz} is among the prime goals of on-going
heavy-ion research at the Relativistic Heavy Ion collider (Beam-Energy
Scan at RHIC) at the Brookhaven National Laboratory (BNL) and one of the
prime motivations for the construction of the Facility for Anti-Proton
and Ion Research (FAIR) in Darmstadt and the Nuclotron-based Ion
Collider Facility (NICA) in Dubna.

The low-energy regime of QCD is governed by approximate chiral symmetry
in the light-quark sector, which is spontaneously broken in the vacuum
via the formation of a quark condensate, $\erw{\bar{q}q} \neq 0$. At
high temperatures and/or densities one expects the restoration of chiral
symmetry, i.e., a cross-over or phase transition.

One problem for theoretical heavy-ion physics is to find possible
signatures of such phase transitions for the highly dynamical medium
created in heavy-ion collisions. From lattice-QCD (lQCD) calculations it
is known that at vanishing net-baryon number, i.e., $\mu_{\text{B}}=0$,
the chiral transition is a rapid cross-over transition at a temperature
of $T_{\text{pc}} \simeq 160 \; \MeV$, while for $\mu_{\text{B}} \neq 0$
lQCD is plagued by the sign problem. Various effective chiral models
like quark-meson $\sigma$- or Nambu-Jona-Lasinio (NJL) models (or
extensions of these with Polyakov loops to implement gluonic degrees of
freedom) indicate that at $\mu_{\text{B}} \neq 0$ the phase transition
becomes of first order, and the first-order phase-transition line in the
phase diagram ends in a critical point, where the phase transition
becomes of second order
 \cite{Kovacs:2006ym,Kovacs:2007sy,Gupta:2011ez,vanHees:2013qla,Wesp:2014xpa,Greiner:2015tra,Kovacs:2016juc}. Possible
indications for this critical point is the divergence of
susceptibilities of conserved quantities like the baryon number or
electric charge and the corresponding grand-canonical fluctuations of
these
quantities \cite{Stephanov:1999zu,Schaefer:2007pw,Skokov:2010uh,Skokov:2010wb,BraunMunzinger:2011ta,Schaefer:2011pn,Morita:2012kt}.

However, the medium created in a relativistic heavy-ion collision
consists of a rapidly expanding and cooling fireball. Thus to find
possible experimental signatures for the different kind of phase
transitions or even to localize the critical point in the QCD phase
diagram a dynamical non-equilibrium treatment is necessary.

On the one hand, at higher collision energies, as achieved at RHIC and
the Large Hadron Collider (LHC) at CERN, the bulk properties of the
medium are astonishingly well described using (viscous) hydrodynamics,
i.e., the fireball is close to local thermal equilibrium. On the other
hand to study fluctuations a kinetic approach is more appropriate. One
intermediate level is to use hydrodynamics and impose fluctuations in
the sense of a Langevin treatment on
top \cite{Nahrgang:2011mg,Nahrgang:2011mv,Herold:2013bi}. Here, however,
usually one assumes Gaussian, white noise for the fluctuations, and the
special statistical properties of a medium close to a phase transition,
particularly around the critical point, where the phase transition
becomes of $2^{\text{nd}}$ order and the correlation lengths are
expected to extent over the entire system, a microscopic treatment in terms
of a Boltzmann-like transport equation is desirable.

In this paper we employ a novel kind of method to derive a numerical
realization of such a kinetic process, based on a linear quark-meson
$\sigma$ model \cite{vanHees:2013qla,Wesp:2014xpa,Greiner:2015tra}. The
particle content of the corresponding quantum field theory are $\sigma$
mesons, pions, and (constituent) quarks. By adjusting the coupling
constants and evaluating the model at finite temperature and density the
different kinds of phase transitions (cross-over, $1^{\text{st}}$, and
$2^{\text{nd}}$ order) can be realized. E.g., at a given value of $g$
with increasing baryochemical potential, $\mu_{\text{B}}$ the order of
the transition (as a function of $T$) can be changed from crossover to
first order through a second-order point. Since the $\sigma$ field is
the order parameter for the chiral phase transition one has to simulate
particles and (mean) fields in a consistent way.

In our approach (Dynamical Solution of a Linear $\sigma$ Model, DSLAM)
we treat the $\sigma$ field as a mean field and the quarks in a
Monte-Carlo test-particle approach to simulate a Vlasov-type of
equation with elastic two-body collisions for the quarks and
anti-quarks. 

To also implement the ``chemical processes'' of $\sigma$-meson decay to
a quark-anti-quark pair and the corresponding reverse process of
quark-anti-quark annihilation, which turn out to be crucial for the
correct realization of the various kinds of phase transitions in the
limit of thermal and chemical equilibrium, we employ the ``wave-particle
duality'' of ``old quantum mechanics''. To this end we realize the mean
field on a spatial grid, which is used to numerically solve for the
mean-field equations. The annihilation process
$q\overline{q} \rightarrow \sigma$ is realized by the test-particle
approach employing the corresponding cross section from the quantum
field theory. Since the $\sigma$ field is realized as a mean field but
not in a particle picture, the energy and momentum of the
$q \overline{q}$ pair is transferred to the mean field in terms of a
relativistic Gaussian wave packet, guaranteeing accurate energy-momentum
conservation. To realize the back reaction, i.e., the decay
$\sigma \rightarrow q \overline{q}$ we use the coarse-graining approach,
i.e., for each grid cell the energy and momentum of the $\sigma$ field
is determined and from this interpreted as a phase-space distribution of
$\sigma$ mesons in local thermal equilibrium employing the equation of
state determined from the quantum-field theoretical model in thermal
equilibrium on the mean-field level. Then in a Monte-Carlo step,
$\sigma$ mesons are decayed to quark-anti-quark pairs, using the
corresponding decay width, consistent with the cross section used for
the pair-annihilation process. In this way again energy-momentum
conservation as well as the principle of detailed balance for the
reactions $\sigma \leftrightarrow q \overline{q}$ is realizable with
high numerical accuracy.

The remainder of the paper is organized as follows. In Sect.\
\ref{sec:2} we present the DSLAM model for the kinetic simulation in
detail (Sect.\ \ref{subsec:2.1}) and demonstrate the validity of
energy-momentum conservation and the stability of the expected
equilibrium solution (Sect.\ \ref{subsec:2.2}) and the correct approach
of a simple off-equilibrium initial state (``thermal quench'') to
thermal and chemical equilibration, particularly the fulfillment of the
principle of detailed balance, in ``box calculations'' (Sect.\
\ref{subsec:2.3}). Finally we simulate an expanding fireball as a simple
model for the kinetics of the medium as created in heavy-ion
collisions (Sect.\ \ref{sec:3}) followed by brief conclusions and
outlook in Sect. \ref{sec:4}.

\section{Non-equilibrium simulation of the linear $\sigma$ model}
\label{sec:2}

\subsection{Particle-field dynamics}
\label{subsec:2.1}

\begin{figure*}
\begin{center}
\includegraphics[width=0.48\linewidth]{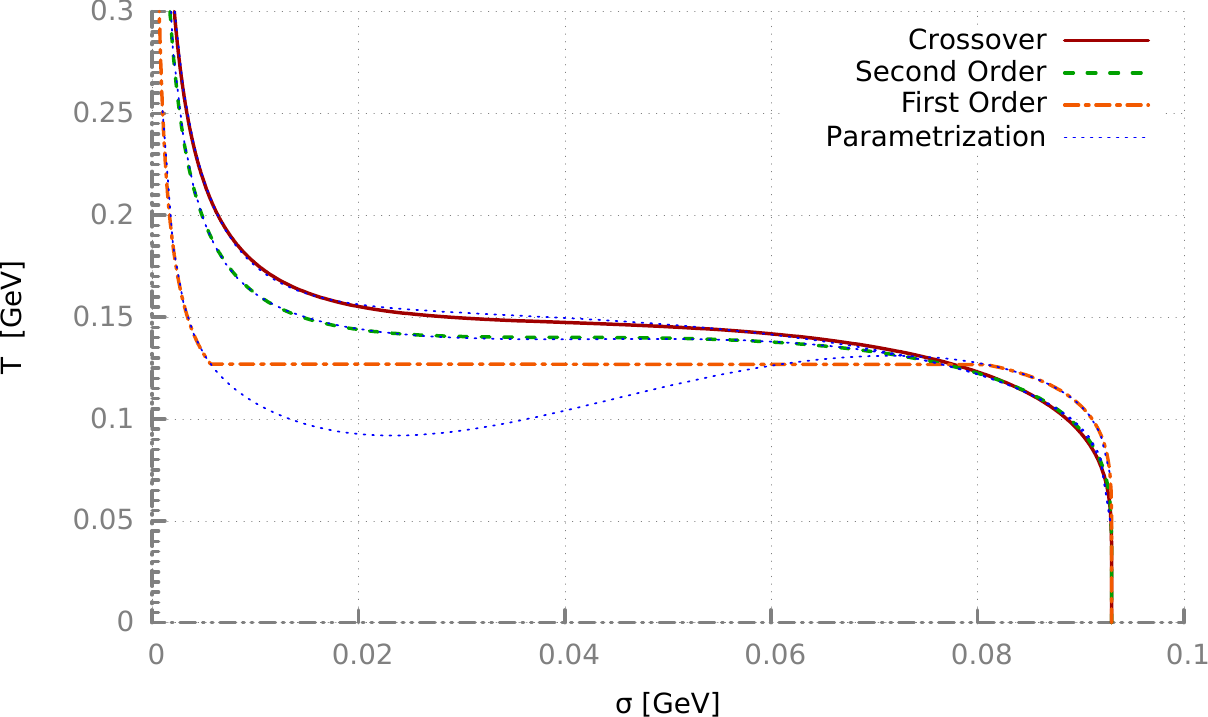}
\includegraphics[width=0.48\linewidth]{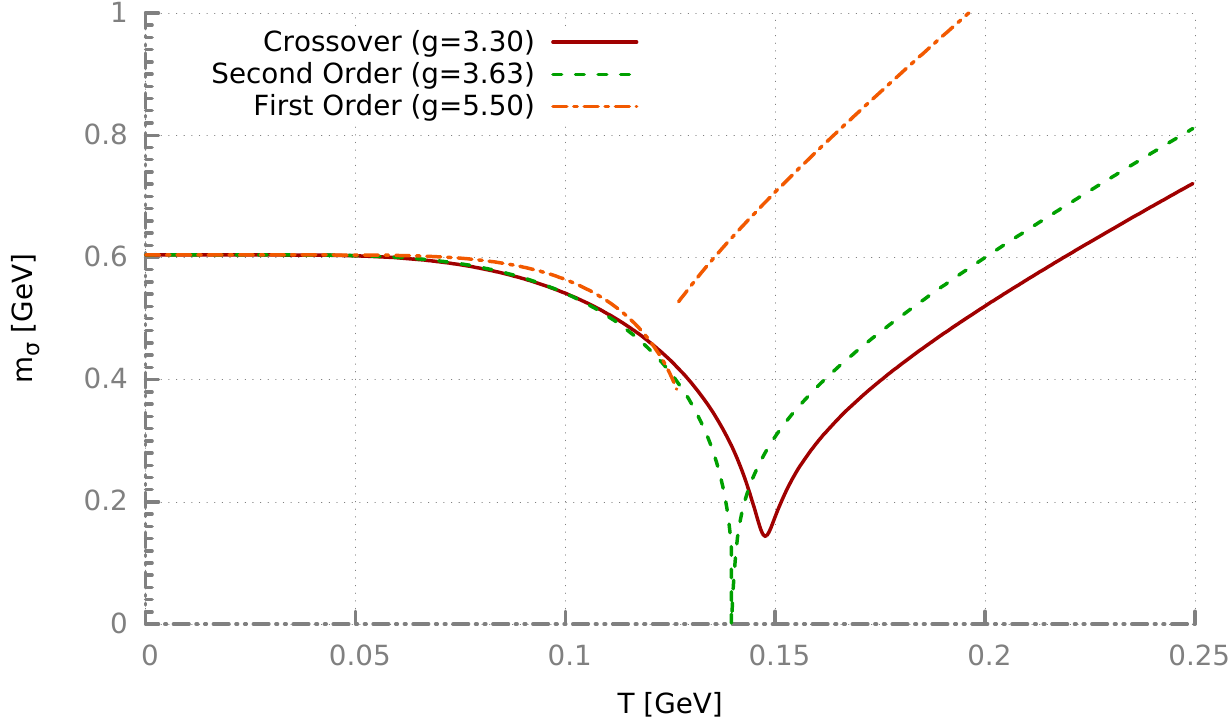}
\end{center}
\caption{\label{fig.1} The phase diagram for the linear $\sigma$ model
  in mean-field approximation at $\mu_{\mathrm{B}}=0$,
  $\erw{\vec{\pi}}=0$. Left: the order parameter $\erw{\sigma}$ and the
  effective $\sigma$ mass
  $m_{\sigma}=\partial^2 \Omega/\partial \sigma^2$.  }
\end{figure*}

In the following we simulate the dynamics of the chiral phase transition
using the linear quark-meson $\sigma$ model based on the chiral symmetry
group $\text{SU}(2)_{\text{L}} \times \text{SU}(2)_{\text{R}}$, acting
separately on the left- and right-handed parts of the two-flavor quark
field $\psi=(u,d)$ and real-valued scalar fields $(\sigma,\vec{\pi})$, which
transform under the SO(4) representation of the chiral group. The
Lagrangian is
\begin{equation}
\begin{split}
\label{2.1}
\mathcal{L}= &\overline{\psi} [\ii \fslash{\partial}-g(\sigma + \ii
\gamma_5 \vec{\pi} \cdot \vec{\tau})] \psi \\
& + \frac{1}{2} (\partial_{\mu}
\sigma \partial^{\mu} \sigma + \partial_{\mu} \vec{\pi}
\cdot \partial^{\mu} \vec{\pi}) - U(\sigma,\vec{\pi}).
\end{split}
\end{equation}
The meson-field potential includes explicit chiral symmetry breaking and
is given by
\begin{equation}
\label{2.2}
U(\sigma,\vec{\pi}) = \frac{\lambda^2}{4} (\sigma^2+\vec{\pi}^2-\nu^2)^2
- f_{\pi} m_{\pi}^2 \sigma-U_0.
\end{equation}
With $\nu^2=f_{\pi}^2-m_{\pi}^2/\lambda$ the minimum of this potential
is given by the vacuum expectation value $\sigma_0=f_{\pi}$, i.e., the
approximate chiral symmetry is broken to the $\mathrm{SU}(2)_{\text{V}}$
isospin symmetry, and the mass of the $\sigma$ mesons is given by
$m_{\sigma}^{(0)} = 2 \lambda^2 f_{\pi}^2+m_{\pi}^2$. Choosing
$\lambda \simeq 20$ leads to $m_{\sigma}^{(0)} \simeq 600 \,\MeV$. Through the
Yukawa couplings of the $\sigma$ field to the quark field in (\ref{2.1})
the quarks acquire the constituent-quark mass $m_{q}=g^2 \sigma_0^2$,
and the variation of $g \in (3.3,5.5)$ leads to a cross-over ($g=3.3$),
a second-order ($g=3.63$), or first-order phase transition at finite
temperature and vanishing baryochemical potential, as determined from
the grand-canonical potential in the mean-field approximation for the
mesons,
\begin{equation}
\label{2.3}
\Omega(T,\mu)=U(\sigma,\vec{\pi}) + \Omega_{\overline{\psi} \psi}
\end{equation}
with the ideal-gas grand-canonical quark potential
\begin{equation}
  \begin{split}
\label{2.4}
\Omega_{\overline{\psi} \psi} = -d_n \int \frac{\dd^3 \vec{p}}{(2
  \pi)^3} [ & T \ln (1+\exp(-\beta(E-\mu))) \\
&+T \ln (1+\exp(-\beta(E+\mu))) ]
\end{split}
\end{equation}
with $E=\sqrt{\vec{p}^2+g^2 (\sigma^2+\vec{\pi}^2)}$ and the degeneracy
factor of the quarks $d_n=2N_f N_c=12$. The mean fields in thermal
equilibrium are determined by the equilibrium condition
$\partial \Omega/\partial \sigma=\partial \Omega/\partial
\vec{\pi}=0$.
In this paper we only consider the case $\erw{\vec{\pi}}=0$. In Fig.\
\ref{fig.1} the results for the mean field $\erw{\sigma}$ for vanishing
baryochemical potential $\mu=0$ and the resulting $\sigma$-meson mass,
$m_{\sigma}=\partial^2 \Omega/\partial \sigma^2$, are shown.
To simulate non-equilibrium simulations we employ our recently developed
semiclassical Monte-Carlo algorithm DSLAM (Dynamical Simulation of a
Linear Sigma Model) \cite{Wesp:2014xpa} to evaluate the dynamics of the
$\sigma$ mean field on the one hand and the quarks as particles in a
Boltzmann-transport approach on the other hand. The challenge here is
that to achieve both thermal and chemical equilibrium besides the usual
collision terms for $qq$ and $q \overline{q}$ elastic scattering one
also has to take ``chemical'' processes like
$q \overline{q} \leftrightarrow \sigma$ into account. Here,
energy-momentum conservation as well as the principle of detailed
balance have to be ensured to guarantee the correct equilibrium limit of
the simulation.

Schematically our approach can be summarized in a stochastic mean-field
and Boltzmann-Vlasov transport \linebreak equation,
\begin{alignat}{2}
\label{2.5}
&\Box \sigma + \lambda(\sigma^2-\nu^2) \sigma - f_{\pi} m_{\pi}^2 + g
\erw{\overline{\psi} \psi} = I(\sigma \leftrightarrow \overline{q} q),
\\
\begin{split}
\label{2.6}
&\left [\partial_t + \frac{p}{E_q} \cdot \vec{\nabla}_{\vec{x}} -
\vec{\nabla}_{\vec{x}} E_{\psi}(t,\vec{x},\vec{p}) \cdot
\vec{\nabla}_{\vec{p}}) \right ] f_{q}(t,\vec{x},\vec{p} ) \\
& \qquad = C(\psi \psi
\rightarrow \psi \psi,\sigma \leftrightarrow \overline{q} q),
\end{split}
\end{alignat}
where $I$ and $C$ denote collision terms contributing in a stochastic
way to the time evolution of the mean field $\sigma$ and the quark
phase-space distribution function $f$, respectively.

This scheme is numerically realized on a space-time grid, and the
elastic binary collisions of the quarks are simulated with the usual
stochastic Monte-Carlo method assuming a constant cross section of
$\sigma_{\text{elastic}}=15 \, \text{mb}$.

The decay process $\sigma \rightarrow \overline{q} q$ and the reverse
recombination process $\overline{q} q \rightarrow \sigma$ are defined by
the quantum field theory as processes involving the ``particle
picture'', leading to (perturbative) decay rates and cross sections in
terms of the corresponding $S$-matrix elements. The corresponding
``interaction processes'' are considered as local, quasi-discrete,
events in comparison to the involved macroscopic scales of the spatial
and temporal changes of the mean field. Thus, within our mean-field
approach to describe the mesons we need to define, how to transfer
energy and momentum from and to the mean field corresponding to the
stochastic and discrete decay and recombination processes.

For the recombination process, the initial state is already given in
terms of the test quarks and anti-quarks, and the cross section is
determined by the decay width of the $\sigma$ meson via the Breit-Wigner
cross section
\begin{equation}
\label{2.7}
\sigma_{q \overline{q} \rightarrow \sigma}(s)=\frac{\Gamma^2}{(\sqrt{s}-m_{\sigma})^2+(\Gamma/2)^2}
\end{equation}
with the decay width
\begin{equation}
\label{2.8}
\Gamma=\frac{g^2}{8 \pi m_{\sigma}} \sqrt{1-\frac{4m_q^2}{m_{\sigma}^2}}.
\end{equation}
Now energy-momentum conservation dictates the kinematics of the
annihilation process, defining the energy and momentum, $\Delta E$ and
$\Delta \bvec{p}$, to be transferred to the classical $\sigma$
field. This is done by adding a Gaussian wave packet, taking into account
the momentum in terms of a Lorentz boosted wave packet at rest. E.g., if
the momentum is in $x$ direction one uses
\begin{equation}
\label{2.9}
\delta \phi(t,\bvec{x}) = A_0 \exp \left [-\frac{\gamma^2(x-v_x t)^2+y^2+z^2}{2
    \sigma^2} \right ]
\end{equation}
with $v_x=p_x/E$ and $\gamma=(1-v_x^2)^{-1/2}$. The width $\sigma$ is a
free parameter of the model and characterizes an interaction volume;
$A_0$ and $\bvec{v}$ have to be determined by the change of the energy
and momentum of the field, which is defined via the usual
energy-momentum tensor,
\begin{alignat}{2}
\label{2.10}
E[\sigma]&=\int_{V} \dd^3 \bvec{x} \left [\frac{1}{2} \dot{\sigma}^2 + \frac{1}{2}
  (\bvec{\nabla} \sigma)^2 + U(\sigma) \right], \\
\label{2.11}
\bvec{P}[\sigma] &= \int_{V} \dd^3 \bvec{x} \dot{\sigma} \bvec{\nabla} \sigma.
\end{alignat}
Then $A_0$ and $\bvec{v}$ are determined such that
\begin{equation}
\label{2.12}
\Delta E=E[\sigma+\delta \sigma]-E[\sigma], \quad \Delta
\bvec{P}=\bvec{P}[\sigma+\delta \sigma]-\bvec{P}[\sigma],
\end{equation}
where $\delta \sigma$ is defined by the Gaussian wave packet (\ref{2.9}).

For the evaluation of the reverse decay process,
$\sigma \rightarrow \overline{q} q$, we have to translate the mean-field
description into a particle phase-space distribution function. To
achieve this, we use the ``coarse-graining'' approach, i.e., we map the
local energy-momentum distribution on the grid to local
thermal-equilibrium distributions, i.e., we parameterize the phase-space
distribution function of $\sigma$ particles by the Maxwell-J{\"u}ttner
distribution,
\begin{equation}
\label{2.13}
f_{\sigma}(x,\bvec{p}) = \exp \left (-\frac{u \cdot p}{T} \right) = \exp
\left[-\frac{\gamma}{T} (E-\bvec{v} \cdot \bvec{p}) \right ],
\end{equation}
where
\begin{equation}
\label{2.14}
\bvec{v}=\frac{\bvec{P}[\sigma]}{E[\sigma]},\quad
\gamma=\frac{1}{\sqrt{1-\bvec{v}^2}}, \quad u=\gamma \vv{1}{\bvec{v}}
\end{equation}
denotes the collective flow-velocity field $\bvec{v}$ and the
four-vector flow field $u$ of the fluid cell $\Delta V$,
respectively. The temperature, $T$, is fixed by demanding
\begin{equation}
\label{2.15}
\epsilon=\frac{E[\sigma]}{\Delta V} = \int_{\R^3} \frac{\mathrm{d}^2 \bvec{p}}{(2
  \pi)^3} \sqrt{m_{\sigma}^2+\bvec{p}^2} f_{\sigma}(x,\bvec{p}).
\end{equation}
Hereby, the $\sigma$ mass and temperature $T$ are determined
self-consistently with the $\sigma$ mass corresponding to the
equilibrium value at this temperature.

Now the phase-space distribution function $f_{\sigma}$ is used in a
Monte-Carlo step to determine a $\sigma$ meson which decays within the
time step according to the decay rate ($\sigma$ width) (\ref{2.8}). The
corresponding energy and momentum of the decayed $\sigma$ meson is taken
out of the $\sigma$ field by again subtracting the appropriate Gaussian
wave packet (\ref{2.9}) in the same way as explained above in connection
with the quark-anti-quark annihilation process, and the quark and
anti-quark with their respective momenta, also determined by the
Monte-Carlo step, are added to the test-particle sample.
\begin{figure*}[t]
\includegraphics[width=0.48\linewidth]{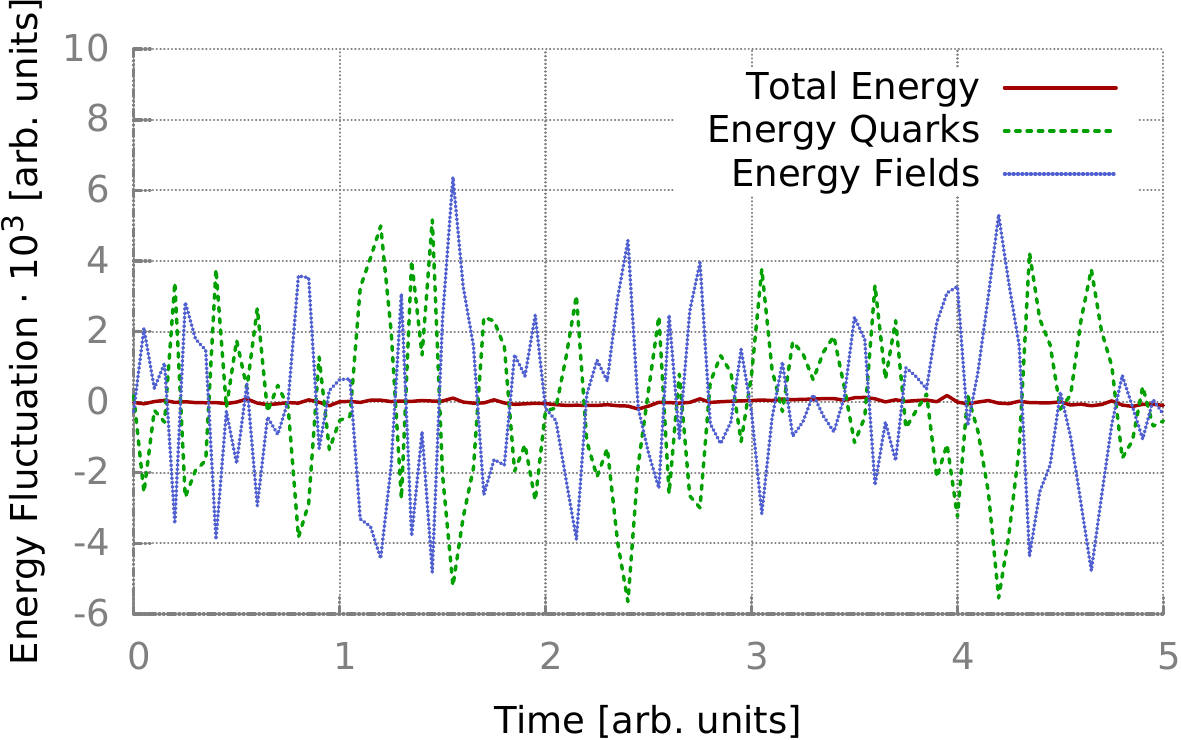}
\includegraphics[width=0.48\linewidth]{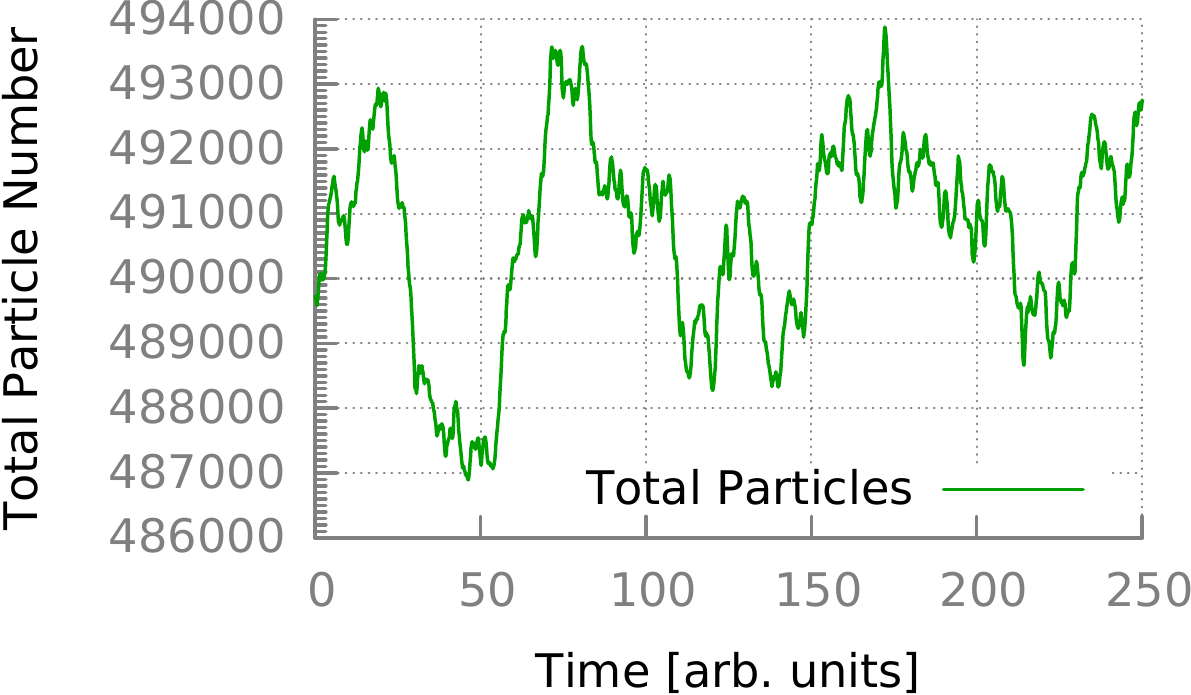}
\caption{(Color online) \textbf{Left:} Energy fluctuations of the mean
  field and the quarks, $\Delta E=E(t)-\erw{E}$. Due to the $\sigma$
  decay and quark-anti-quark annihilation processes energy is exchanged
  between the mean field and the particles, leading to thermal
  fluctuations $|\Delta E|/E \sim 10^{-2}$ for the field and
  $|\Delta E|/E \sim 10^{-3}$ for the quarks and anti-quarks. The total
  energy is conserved up to numerical fluctuations of
  $|\Delta E_{\text{tot}}/\erw{E} \lesssim 5 \cdot
  10^{-5}$. \textbf{Right:} Fluctuations of the total quark number due
  to $\sigma$-decay and $q \bar{q}$-annihilation processes.}
\label{fig.2}
\end{figure*}

\subsection{Test calculations in equilibrium}
\label{subsec:2.2}

To validate the above defined algorithm to simulate the kinetics of the
linear $\sigma$ model in our particle-field approach, we have performed
some numerical tests by simulating a microcanonical ensemble in a
finite-size cubic box employing periodic boundary conditions. The number
of test particles is $N_{\text{test}}=3 \cdot 10^6$, the numerical time
steps $\Delta t=2 \cdot 10^{-3} \text{a.u.}$ (a.u.: arbitrary units),
the spatial grid for the field is $N_{\sigma}=128^3$, corresponding to a
total volume $V=(6 \;\fm)^3$. The total simulation time is
$300 \; \text{a.u.}$ corresponding to $150000$ time steps. The
interaction volume of the Gaussian parameterization of the wave packets
(\ref{2.9}) is $32^3$ grid points, i.e., approximately $1.5\%$ of the
system volume. As an initial condition we have chosen equilibrium
conditions at a given temperature.

In Fig.\ \ref{fig.2} we show the fluctuations of the field and quark-anti-quark
energy (left) and the total quark number (right) due to the
$\sigma$-decay and $q\bar{q}$ annihilation processes, which show thermal
fluctuations around the equilibrium mean values, while the total energy
is conserved with a numerical accuracy of
$|\Delta E_{\text{tot}}/\erw{E} \lesssim 5 \cdot 10^{-5}$.
\begin{figure*}[t]
\includegraphics[width=0.48\linewidth]{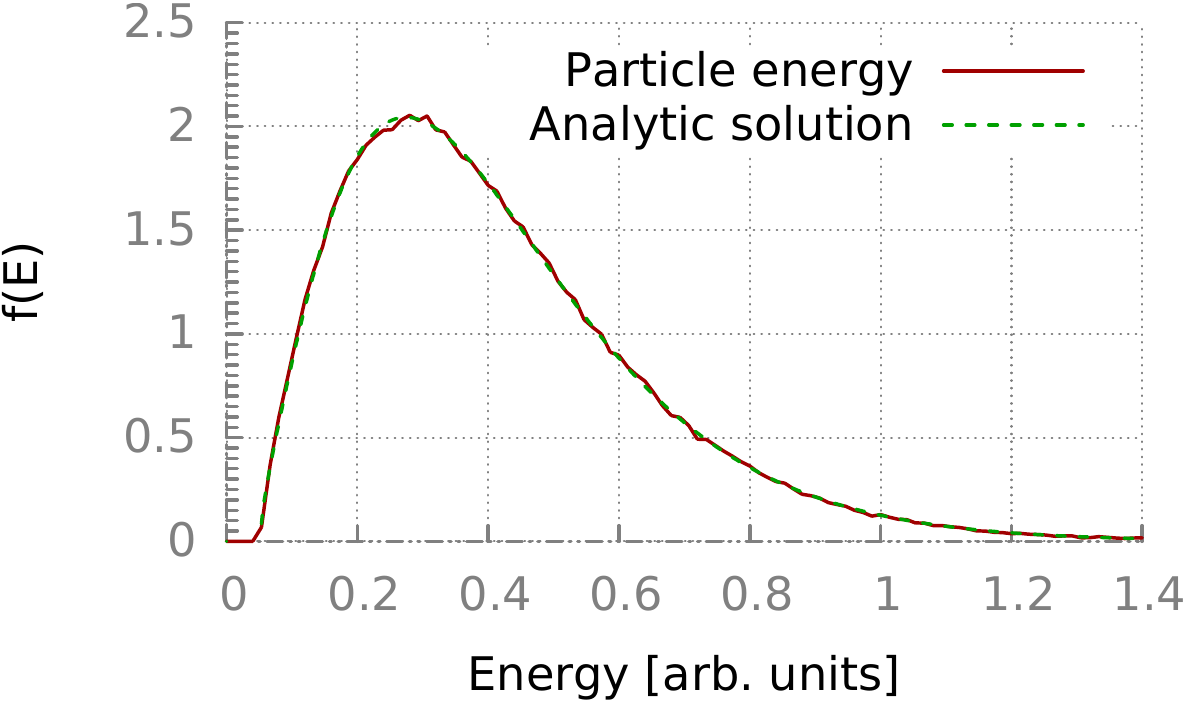}
\includegraphics[width=0.48\linewidth]{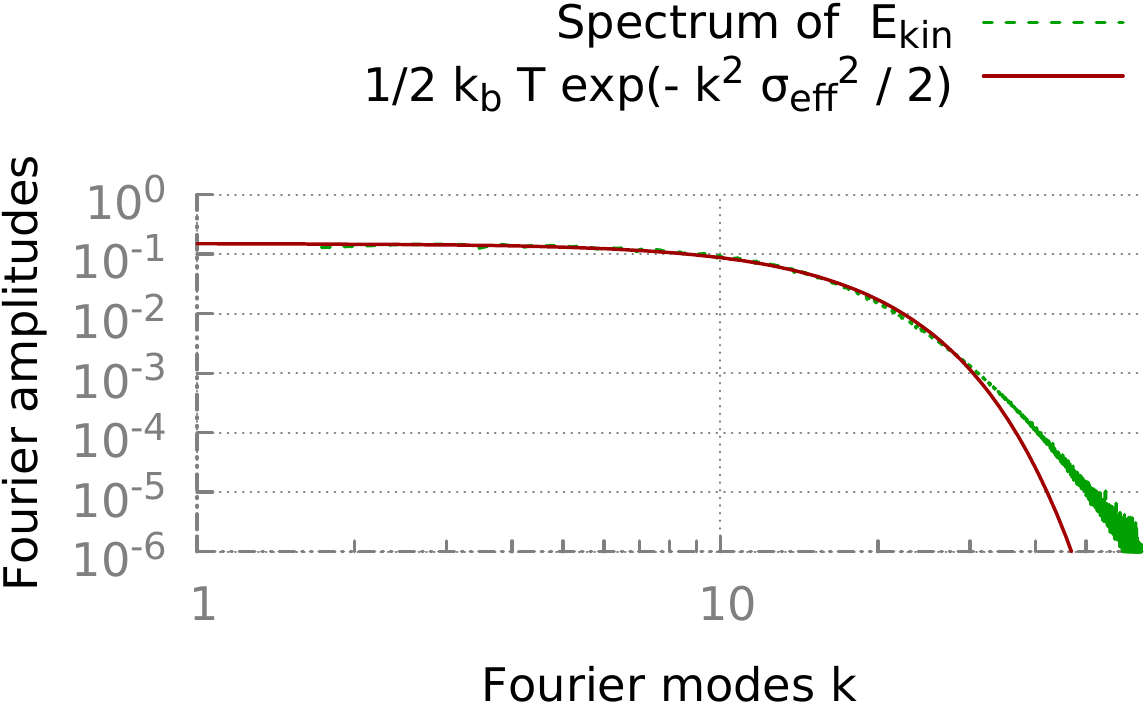}
\caption{(Color online) \textbf{Left:} The energy distribution of the
  quarks and anti-quarks follow the expected relativistic
  Maxwell-Boltzmann distribution $f_q(E) \propto \exp(-E/T)$,
  demonstrating the numerical stability of the kinetic equations as well
  as the validity of the principle of detailed balance. \textbf{Right:}
  The spectrum of the kinetic energy density of the mean field,
  $\dot{\sigma}^2/2$. For low $k$ (long wavelengths) the distribution
  follows the expected equipartion theorem for a classical field,
  according to which each mode carries a mean energy of $1/2 kT$, while
  at higher energies the finite interaction volume encoded in the finite
  width of the Gaussian wave packets leads to an effective cut-off
  $\sigma_{\text{eff}}^2=\sigma_x^2+\sigma_y^2+\sigma_z^2=3 \sigma^2$.}
\label{fig.3}
\end{figure*}

As can be seen in Fig.\ \ref{fig.3}, the energy distribution of the
quarks and anti-quarks (left panel) follows the expected relativistic
Maxwell-Boltzmann distribution
\begin{equation}
\label{2.16}
\frac{\dd N}{\dd E \dd^3 \bvec{x}} = \frac{1}{2 \pi^2} E \sqrt{E^2-m_q^2}
\exp \left (-\frac{E}{T} \right)
\end{equation}
with high numerical accuracy, which demonstrates again the good
fulfillment of energy conservation as well as the principle of detailed
balance for the elastic quark/anti-quark scattering and the ``chemical''
processes $q \overline{q} \leftrightarrow \sigma$. In the right panel we
show the spectrum of the kinetic energy of the mean field,
$\dot{\sigma}^2/2$. From classical field theory one expects a constant
average energy of $k_{\text{B}} T$ for each mode, which is indeed the
case for small wave number $k$, i.e., long wave lengths
$\lambda=2 \pi/k$. At lower wave lengths the finite interaction volume
encoded in the finite width of the Gaussian wave packets (\ref{2.9})
used to describe the energy-momentum transfer in the interaction between
quarks/anti-quarks and $\sigma$ mesons from and to the mean field, leads
to an effective cut-off of
$1/\sigma_{\text{eff}}^2=\sigma_{x}^2+\sigma_y^2+\sigma_z^2=3 \sigma^2$,
i.e., the power spectrum is expected to follow \cite{Wesp:2014xpa}
\begin{equation}
\label{2.17}
\frac{\dd E_{\sigma}}{\dd k} = \frac{1}{2} k_{\text{B}} T \exp \left (-k^2
\sigma_{\text{eff}}^2/2 \right).
\end{equation}
The deviations from this result at very high modes can be explained by
the non-linear potential in the mean-field equations of motion.
\begin{figure*}[t]
\includegraphics[width=0.48\linewidth]{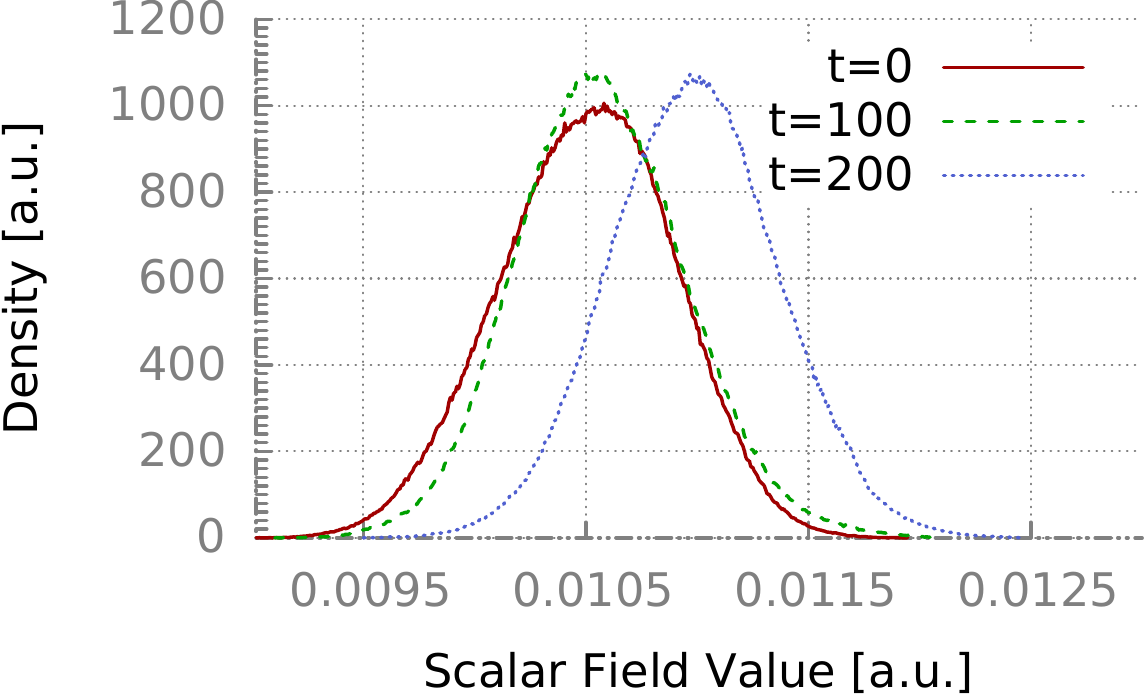}
\includegraphics[width=0.48\linewidth]{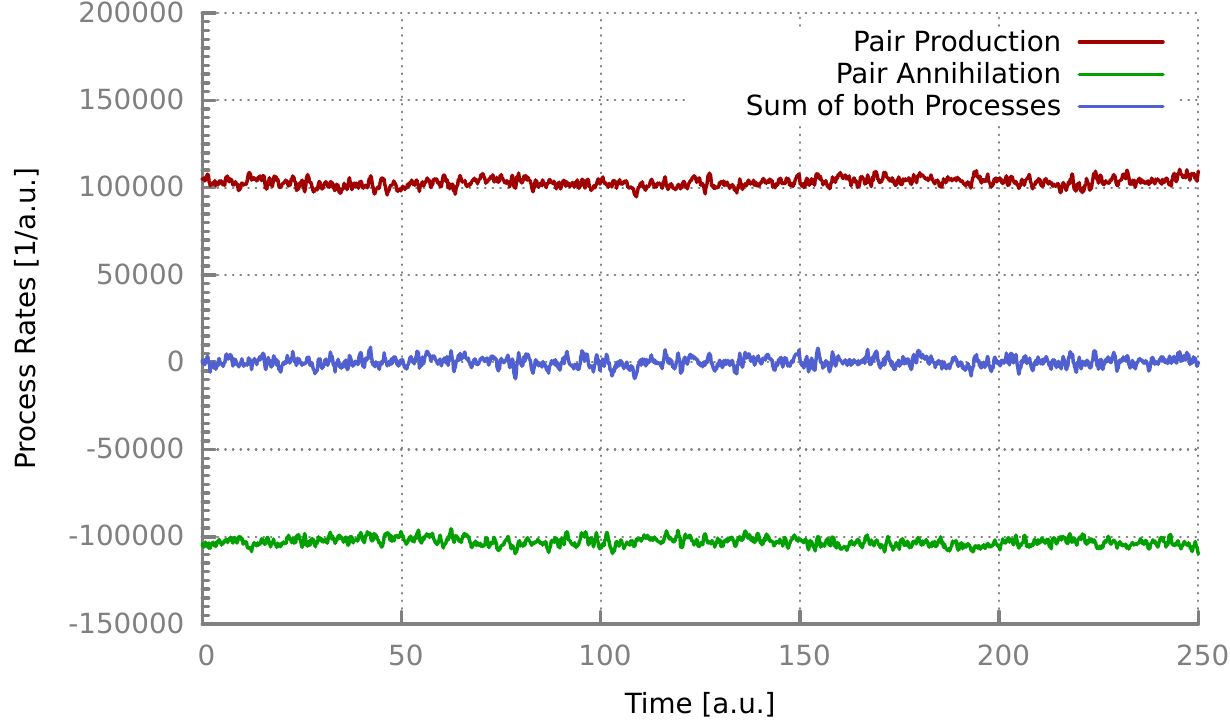}
\caption{(Color online) \textbf{Left:} The distribution of the scalar
  field around its equilibrium value shows the expected Gaussian
  shape. The mean field can drift slowly with time due to the
  particle-field interactions. The quark-annihilation processes increase
  the local fluctuations, while particle production damps them by
  dissipating energy from the field. \textbf{Right:} Volume integrated
  rates for the pair-production and pair-annihilation processes,
  $\sigma \leftrightarrow \overline{q}q$, demonstrating the good
  fulfilment of the principle of detailed balance.  }
\label{fig.4}
\end{figure*}

In Fig.\ \ref{fig.4} (left) we show the distribution of the scalar-field
values around their mean as a function of time. Due to the
quark-antiquark annihilation and $\sigma$-decay processes the mean-field
value can drift slowly, while the fluctuations lead to the expected
Gaussian distribution around this average. The local fluctuations of the
field are increased by $q \overline{q}$ annihilation and damped by
$\sigma$ decay. On the right-hand side the corresponding
volume-integrated rates are plotted, demonstrating the good fulfillment
of the principle of detailed balance within our simulation, which
guarantees the right chemical equilibrium in the long-time limit.

\begin{figure*}[t]
\includegraphics[width=0.48\linewidth]{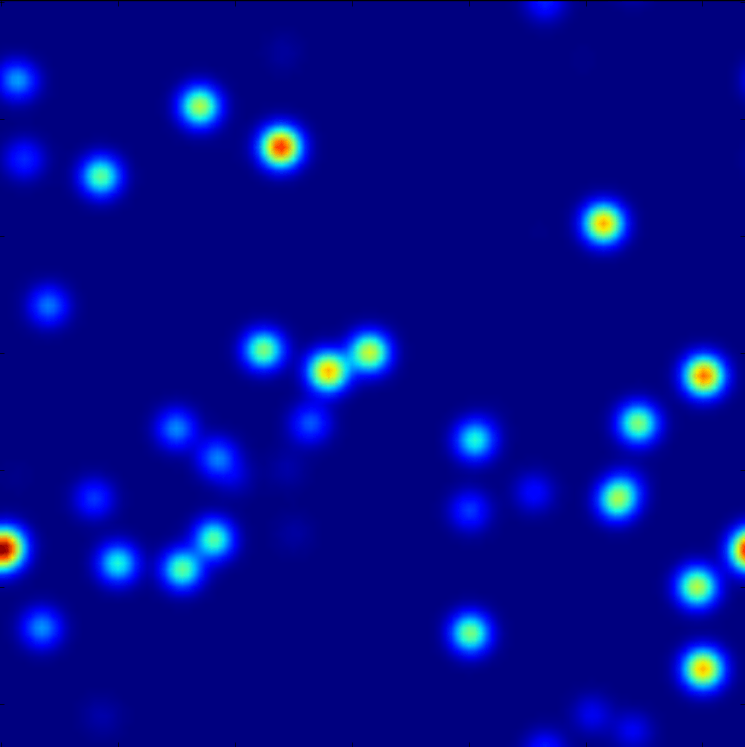}
\includegraphics[width=0.48\linewidth]{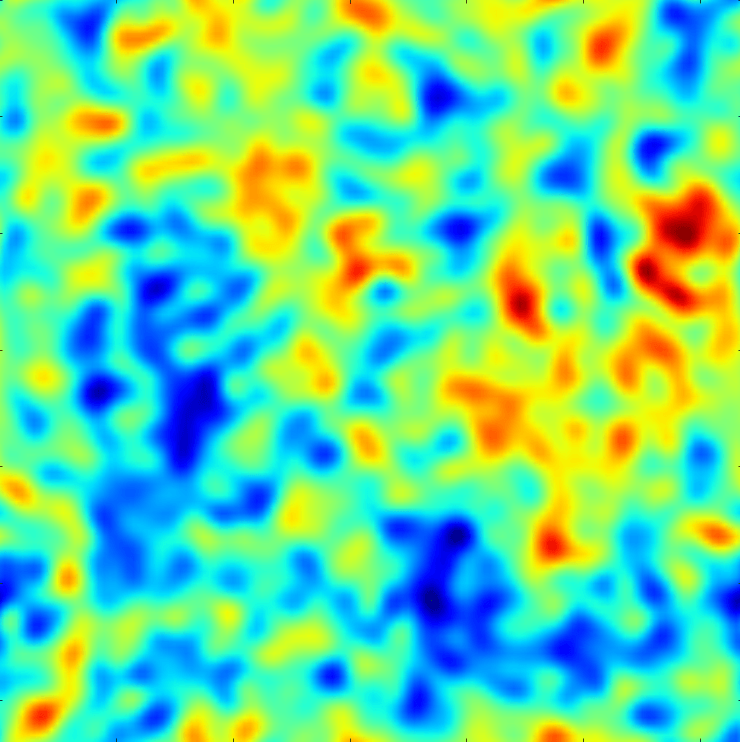}
\caption{(Color online) Contour plot of the scalar field in an
  $x$-$y$-plane cut. The field is initialized with a uniform scalar
  field without excitations. Due to $q\overline{q}$ annihilation, field
  fluctuations build up, leading to single Gaussian excitations around the
  respective random interaction regions at early times
  (\textbf{left}). The \textbf{right} panel shows the field fluctuations
  after chemical equilibrium has been reached, leading to the expected
  thermal fluctuations of the field with height and strength scaling
  with temperature.
}
\label{fig.5}
\end{figure*}

In Fig.\ \ref{fig.5} we demonstrate the chemical-equilibration process
with a contour plot of the scalar field in an $x$-$y$-plane
cut. Initially we have chosen a uniform field without any
excitations. At early times (left panel) one can see the excitations of
the field due to $q\overline{q}$ annihilation in terms of Gaussian blobs
around the random interaction regions. At later times, when chemical
equilibrium is reached (right panel), i.e., the particle-annihilation
and creation rates become equal, we observe the expected thermal
fluctuations. Their spatial width is determined by the interaction
volume governed by the standard deviation, $\sigma$, parameterizing the
interaction volume in the Gaussian wave packets (\ref{2.9}). The height
and strength of the fluctuations in the equilibrium limit is governed by
the temperature $T$ of the system in accordance with the above discussed
power spectrum (right panel of Fig.\ \ref{fig.3}).

\subsection{Thermal quench}
\label{subsec:2.3}

As another test of the field-particle algorithm we initialize the
kinetic equations in an off-equilibrium ``thermal quench'' situation in
an isotropic periodic box. Initially the quarks are sampled at a
temperature $T_q=140 \; \MeV$ and the $\sigma$ field at its equilibrium
value for $T_{\sigma}=180 \; \MeV$. The coupling is set to $g=3.3$
corresponding to a cross-over chiral transition.The
$q \overline{q}$-annihilation cross section is determined with the
corresponding Breit-Wigner cross section as described
above. Additionally a constant elastic $qq \rightarrow qq$-scattering
cross section, $\sigma_{\text{elastic}}=15\;\mathrm{mb}$ is employed. We
use a number of test particles $N_{\text{test}}=2 \cdot 10^6$ and a
time-step size of $\Delta t=0.002 \; \fm/c$.

\begin{figure*}
   \centering
   \begin{subfigure}[t]{0.49\textwidth}
       \centering
       \includegraphics[keepaspectratio=True,width=\textwidth,height=0.2\textheight]{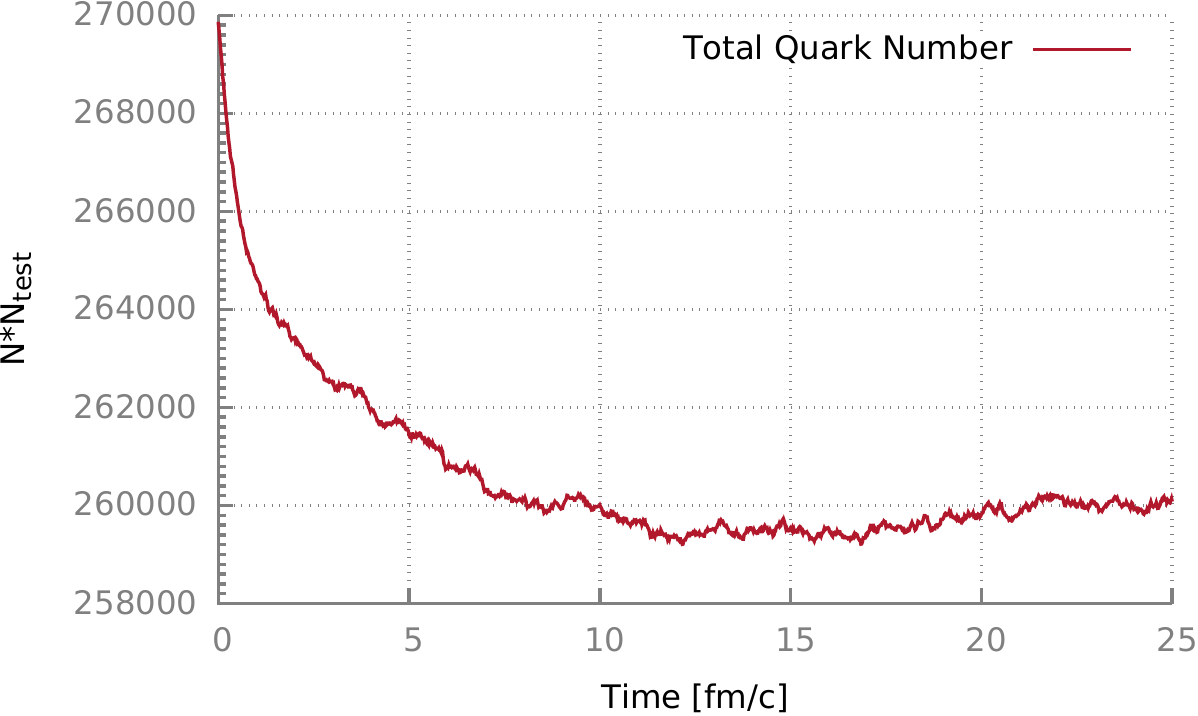}
       \caption{Total number of quarks in the system.}
       \label{fig.6.a}
   \end{subfigure}
   \hfill
   \begin{subfigure}[t]{0.49\textwidth}
       \centering
       \includegraphics[keepaspectratio=True,width=\textwidth,height=0.2\textheight]{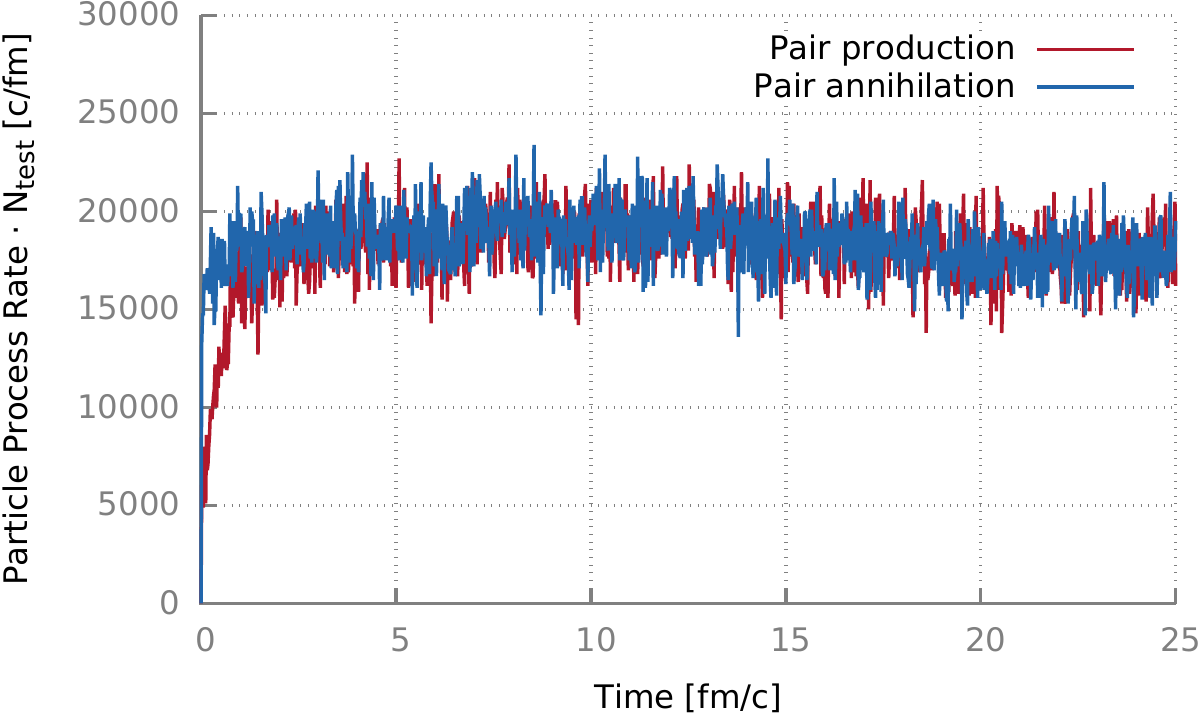}
       \caption{Reaction rate for pair-production and pair-annihilation.}
       \label{fig.6.b}
   \end{subfigure}
\par\bigskip
   \begin{subfigure}[t]{0.49\textwidth}
       \centering
       \includegraphics[keepaspectratio=True,width=\textwidth,height=0.2\textheight]{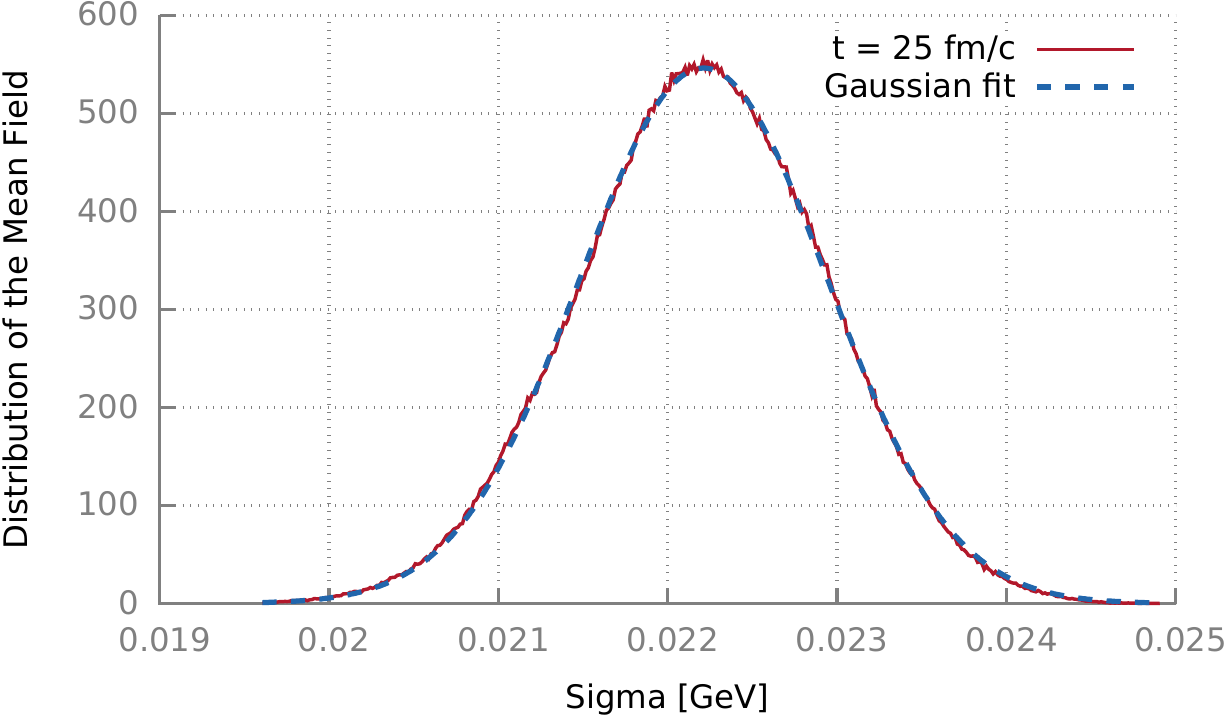}
       \caption{Spatial distribution of the sigma mean field $\langle \sigma(\bvec x) \rangle$ after thermalization time $t=25 \ \textrm{fm}/\textrm{c}$.}
       \label{fig.6.c}
   \end{subfigure}
   \hfill
   \begin{subfigure}[t]{0.49\textwidth}
       \centering
       \includegraphics[keepaspectratio=True,width=\textwidth,height=0.2\textheight]{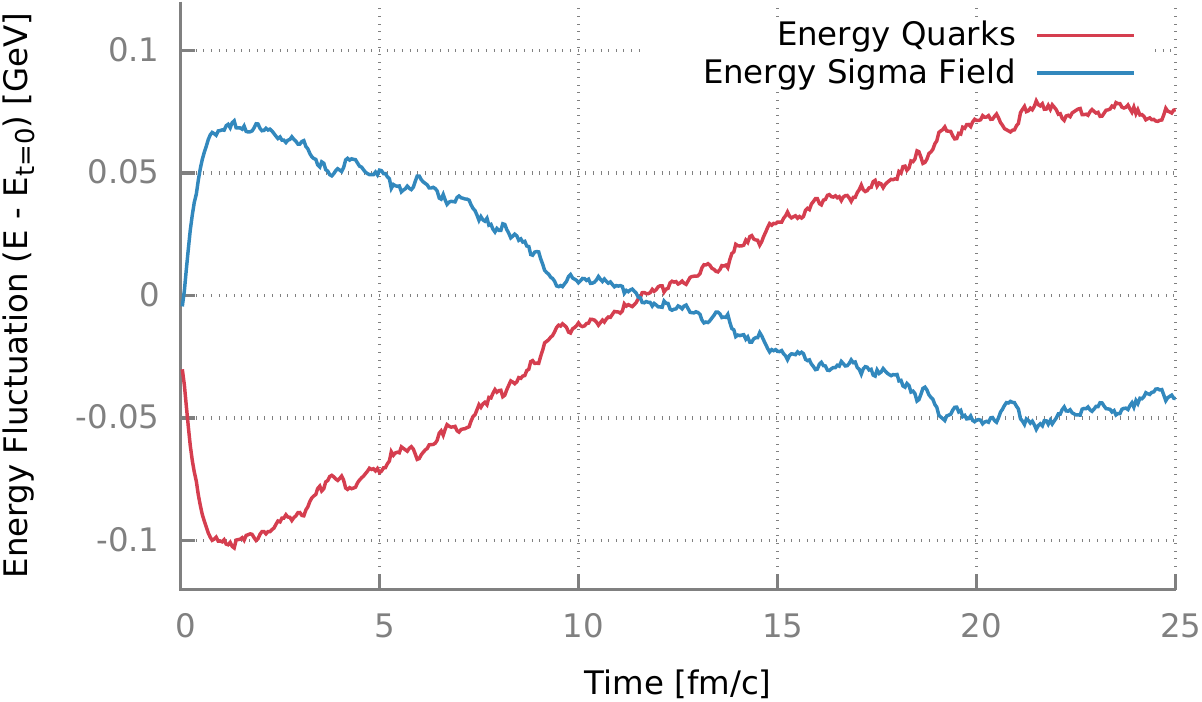}
       \caption{Deviation from the initial energy for the quarks and the energy of the field, calculated with $E(t) - \langle E(t=0) \rangle$.}
       \label{fig.6.d}
   \end{subfigure}
\par\bigskip
   \begin{subfigure}[t]{0.49\textwidth}
       \centering
       \includegraphics[keepaspectratio=True,width=\textwidth,height=0.2\textheight]{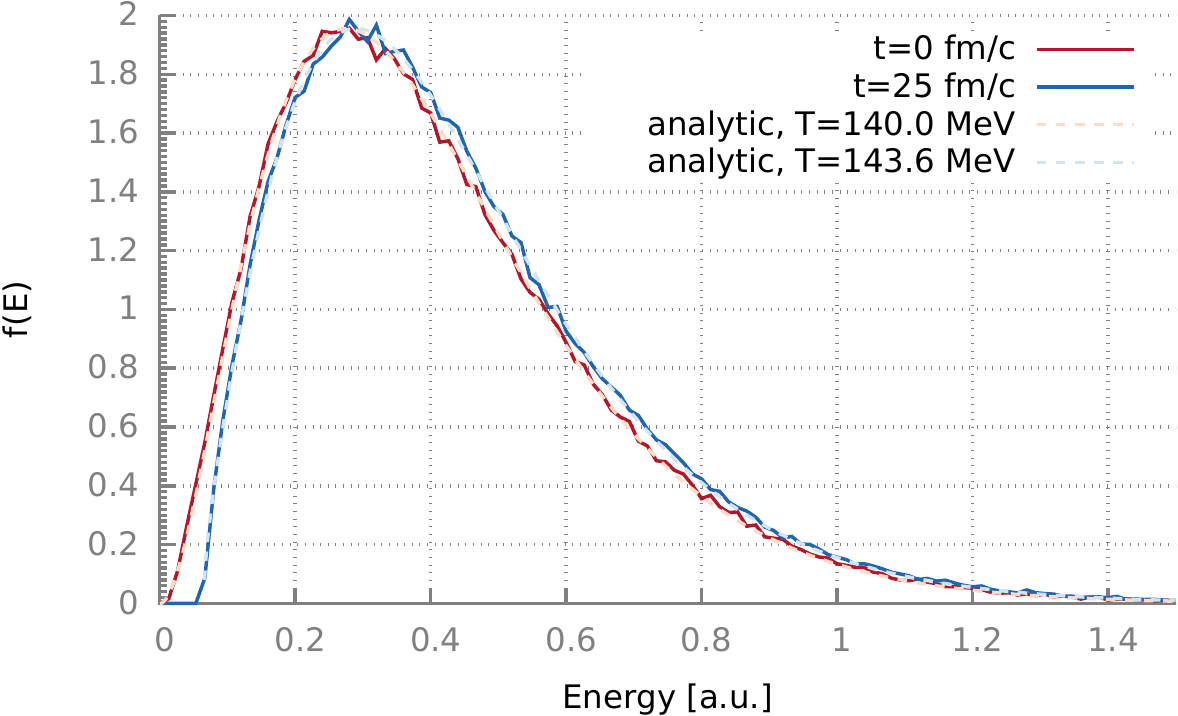}
       \caption{Distribution of the quarks at the beginning of the system and after global thermalization.}
       \label{fig.6.e}
   \end{subfigure}
   \hfill
   \begin{subfigure}[t]{0.49\textwidth}
       \centering
       \includegraphics[keepaspectratio=True,width=\textwidth,height=0.2\textheight]{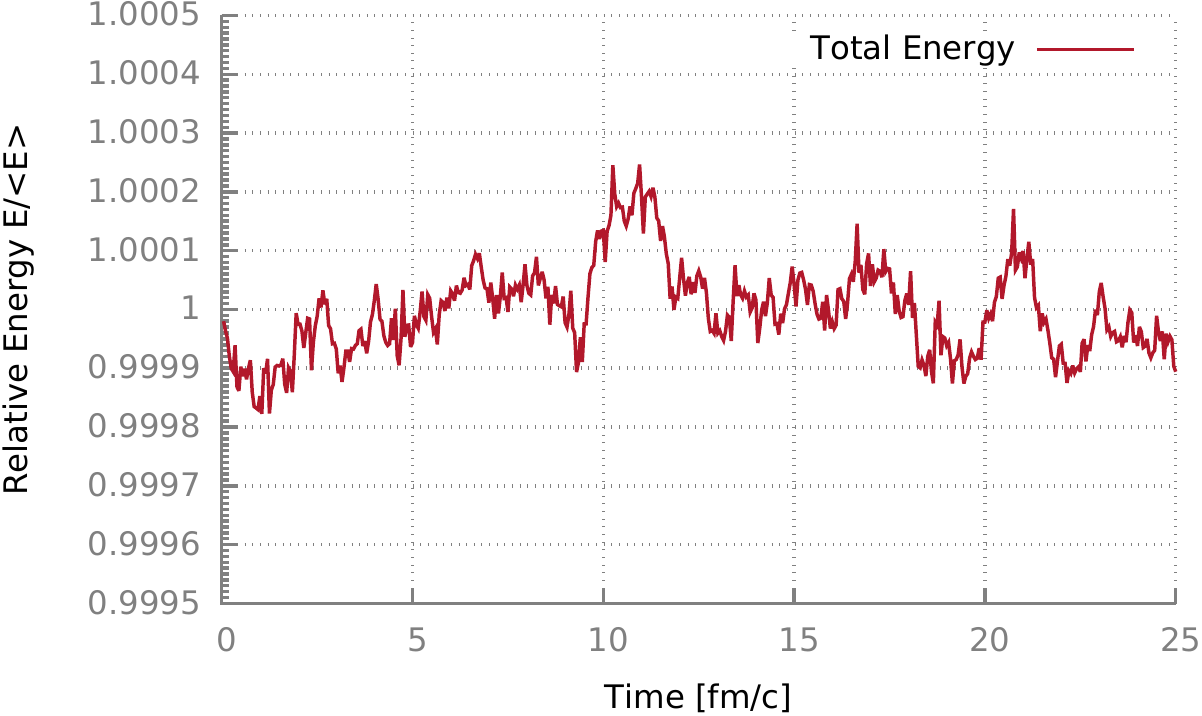}
       \caption{Relative fluctuations of the system's total energy $E(t)/\langle E(t=0) \rangle$.}
       \label{fig.6.f}
   \end{subfigure}
   \caption{\label{fig.6} (Color online) Results for a calculation with the
     quenched-scenario. Particles and fields can interact and their
     interaction rates converge after some time, leading the same global
     temperature of $T_{\text{eq}} \simeq 144 \;\MeV$. The temperature
     of the quarks is given by the Boltzmann-distribution, the
     temperature of the field is given by the distribution of the field
     modes as expected from a power spectrum $S(k) =T/k^2$ as in a
     Langevin approach.}
\end{figure*}
As shown in Fig. \ref{fig.6.a} the total quark number, i.e., the number
of quark-antiquark pairs, drops rapidly in the beginning due to the
annihilation and creation processes
$q \overline q \leftrightarrow \sigma$, which (according to Fig.\
\ref{fig.6.b}) converge after about $10 \; \fm/c$, fluctuating around
the common mean-field value. Thermal equilibrium is roughly reached
after about $20 \;\fm/c$, leading to a common global temperature of
about $T_{\text{eq}} \simeq 144 \;\MeV$ as demonstrated by Figs.\
\ref{fig.6.c} and \ref{fig.6.e}: at $t=25 \; \fm/c$ the quark
distribution is given by a Boltzmann distribution at this temperature,
and the $\sigma$ field shows a Gaussian distribution around its
equilibrium mean value. The temperature change of the quarks due to the
transfer of initial potential field energy by only about $4 \;\MeV$ is
small, because the field's initial potential energy is much smaller as
compared to the energy of the quarks. As one can see in Figs.\
\ref{fig.6.d} and \ref{fig.6.f}, energy between the quarks and the field
is exchanged, while the total energy is conserved with high accuracy. As
can be seen from Fig.\ \ref{fig.6.a} the particle number still rises
slowly, and the system has not fully reached equilibrium. Also the
mean-field value is not exactly at the equilibrium value shown in Fig.\
\ref{fig.1}. The final approach to exact chemical equilibrium is very
slow since the difference in the $\sigma$ potential is very small in
this temperature range around the cross-over transition, resulting in a
weak driving force. This underlines the importance of the ``chemical''
processes $q \overline{q} \leftrightarrow \sigma$ for a description of
the global thermalization of the system to the expected global
equilibrium as described by the phase diagram shown in Fig.\
\ref{fig.1}.

\begin{figure*}
   \centering
   \begin{subfigure}[t]{0.49\textwidth}
       \centering
       \includegraphics[keepaspectratio=True,width=\textwidth,height=0.2\textheight]{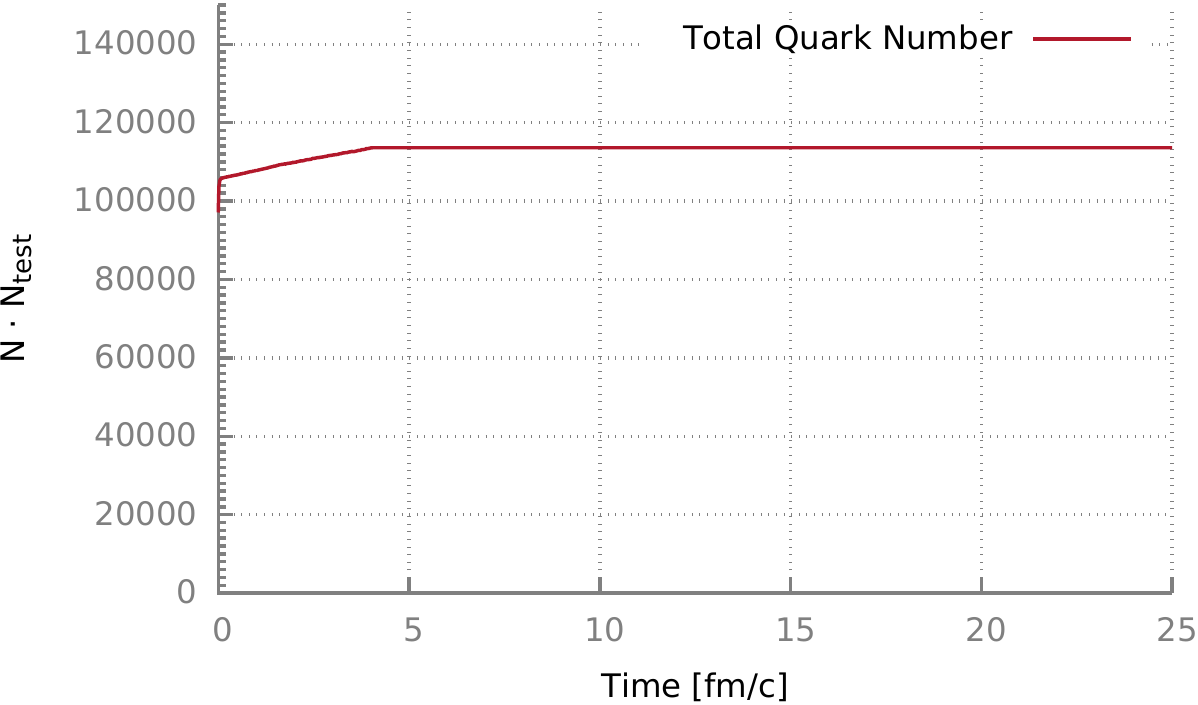}
       \caption{Total number of quarks in the system.}
       \label{fig.7.a}
   \end{subfigure}
   \hfill
   \begin{subfigure}[t]{0.49\textwidth}
       \centering
       \includegraphics[keepaspectratio=True,width=\textwidth,height=0.2\textheight]{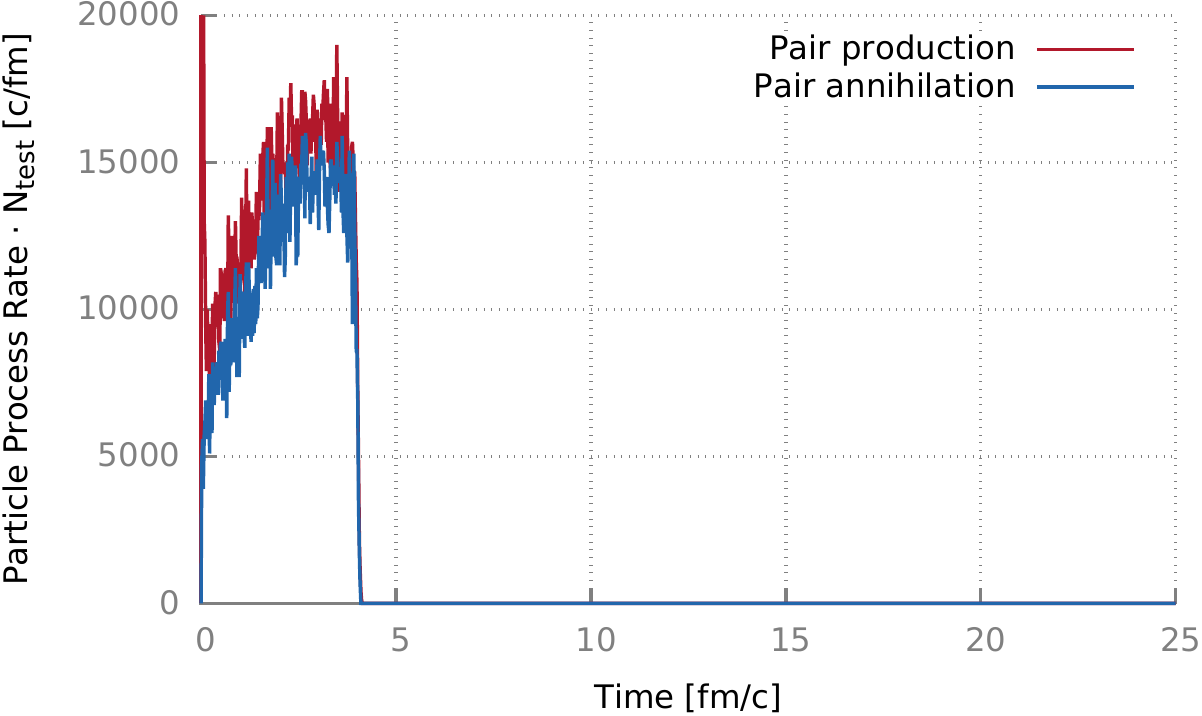}
       \caption{Reaction rate for pair-production and pair-annihilation.}
       \label{fig.7.b}
   \end{subfigure}
\par\bigskip
   \begin{subfigure}[t]{0.49\textwidth}
       \centering
       \includegraphics[keepaspectratio=True,width=\textwidth,height=0.2\textheight]{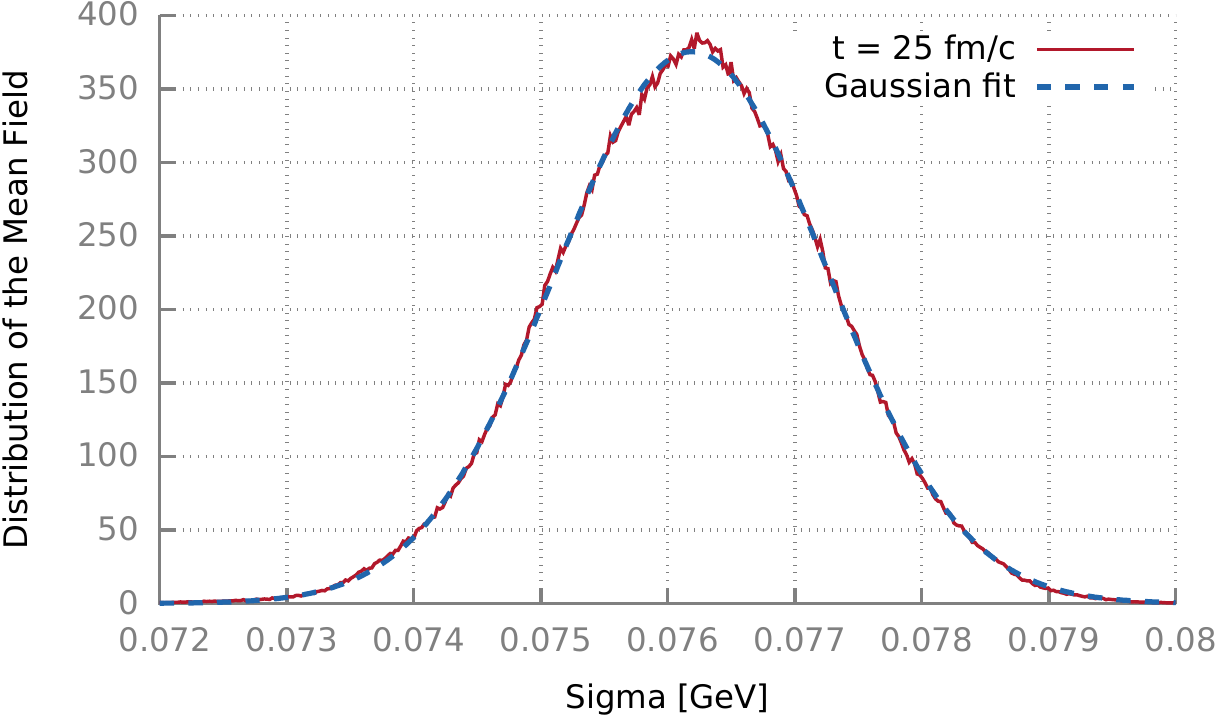}
       \caption{Spatial distribution of the mean $\sigma$ field
         $\langle \sigma(\bvec x) \rangle$ in the stationary
         non-equilibrium state at $t=25 \ \textrm{fm}/c$.}
       \label{fig.7.c}
   \end{subfigure}
   \hfill
   \begin{subfigure}[t]{0.49\textwidth}
       \centering
       \includegraphics[keepaspectratio=True,width=\textwidth,height=0.2\textheight]{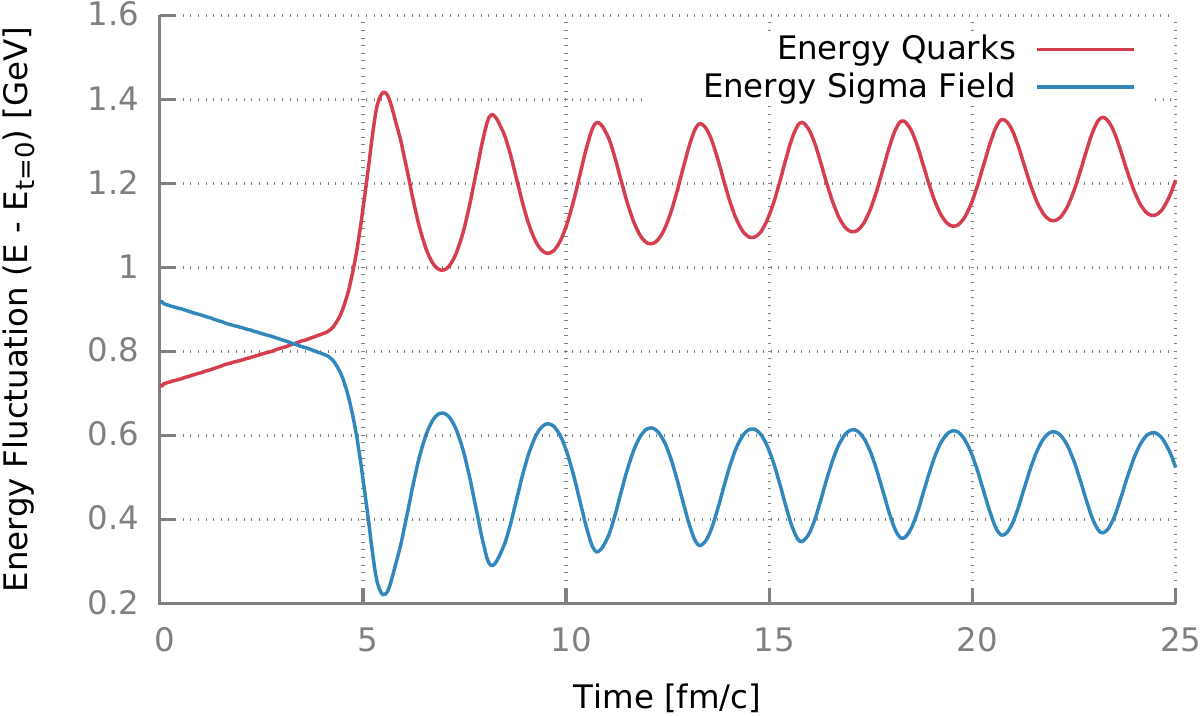}
       \caption{Deviation from the initial energy for the quarks and the
         energy of the field, calculated with
         $E(t) - \langle E(t=0) \rangle$.}
       \label{fig.7.d}
   \end{subfigure}
\par\bigskip
   \begin{subfigure}[t]{0.49\textwidth}
       \centering
       \includegraphics[keepaspectratio=True,width=\textwidth,height=0.2\textheight]{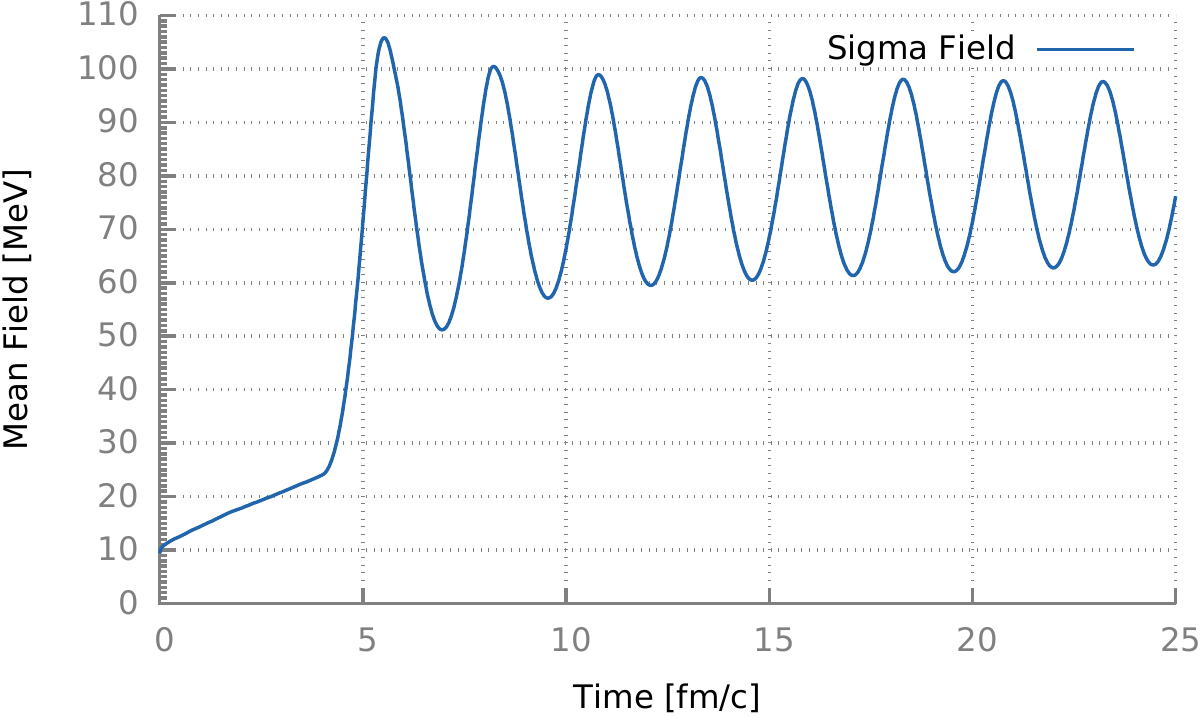}
       \caption{Volume averaged value of the $\sigma$-mean field.}
       \label{fig.7.e}
   \end{subfigure}
   \hfill
   \begin{subfigure}[t]{0.49\textwidth}
       \centering
       \includegraphics[keepaspectratio=True,width=\textwidth,height=0.2\textheight]{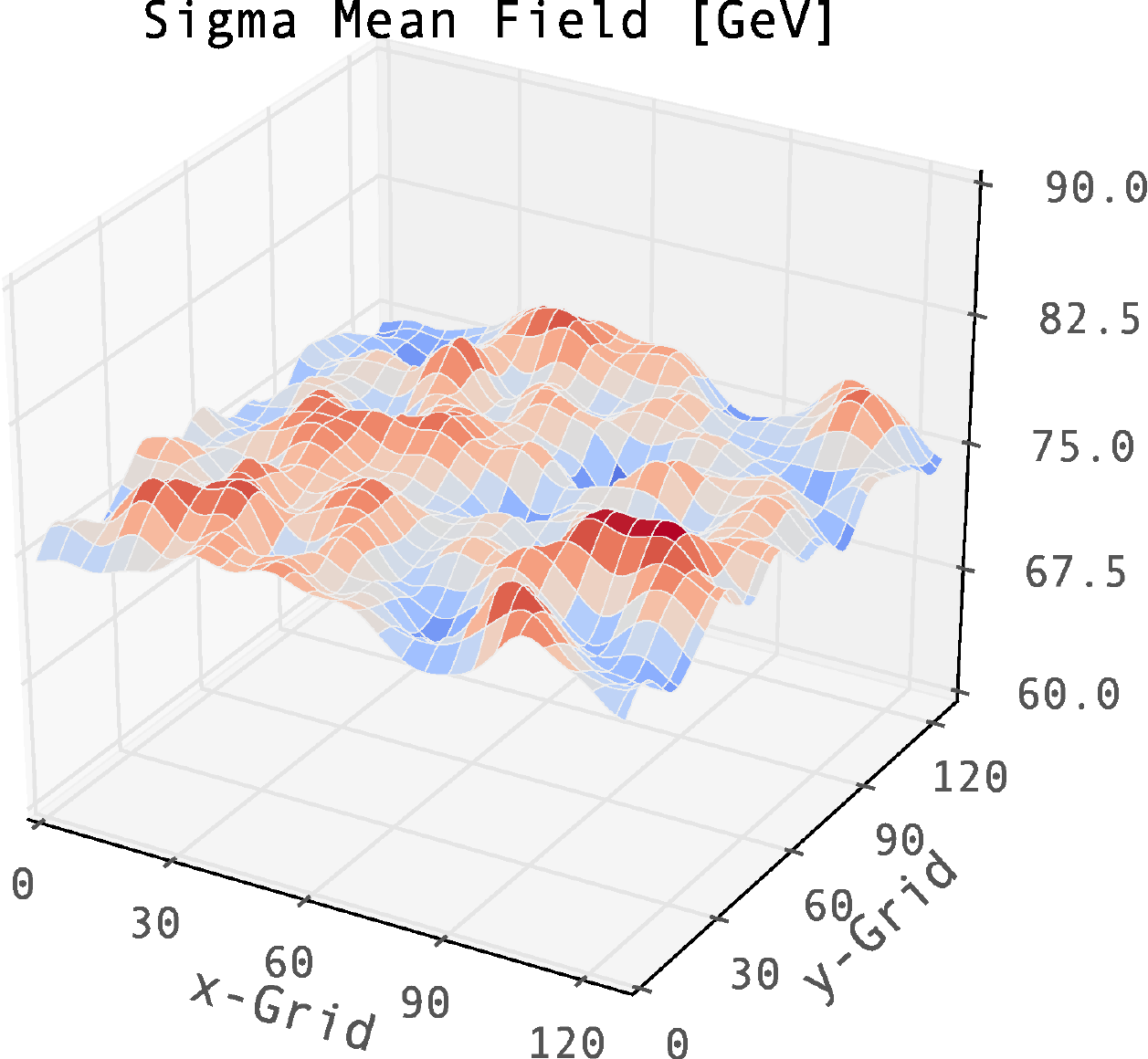}
       \caption{\label{fig:calcs:physical:quench2:all}
       Spatial distribution of the $\sigma$-field at $t=25 \ \textrm{fm}/\textrm{c}$.}
       \label{fig.7.f}
   \end{subfigure}
   \caption{\label{fig.7} (Color online) Results for a calculation with
     the quenched-scenario from the initially hot phase to the cold and
     chiral broken phase. Fields and particles can interact above $T_c$
     and the global shift of the $\sigma$-field is damped by particle
     production. At and below $T_c$ field and particles decouple because
     the $\sigma$-particle can not decay anymore. The system becomes
     undamped and falls in a coherent oscillation.}
\end{figure*}

In the next example we choose the same parameters as in the previous one
but start with a lower Temperature of $T_q=80\;\MeV$ for the quarks. The
$\sigma$ field has been initialized at the value
$\erw{\sigma} \simeq 10 \; \MeV$ in the chiral restored phase. Initially
the quark number raises slightly due to $\sigma$-decay processes as
shown in Figs.\ \ref{fig.7.a} and \ref{fig.7.b} while the mean field
also increases moving towards the cross-over transition to the chiral
broken phase. At some point of the evolution the $\sigma$ mass drops
below the threshold for the decay to quark-anti-quark pairs,
$m_{\sigma}<2 m_q$, and the field becomes undamped and thus keeps
oscillating for the rest of the time evolution (Fig.\
\ref{fig.7.e}). This non-equilibrium state stays in this oscillating
state after about $10 \;\fm/c$. The field shows again a Gaussian
distribution around its average (Fig.\ \ref{fig.7.c}), and energy is
exchanged with the quarks (Fig.\ \ref{fig.7.d}) via the Vlasov terms in
Eqs.\ (\ref{2.5}) and (\ref{2.6}). In Fig. \ref{fig.7.f} one sees that
the spatial distribution of the $\sigma$ field shows the corresponding
oscillations of the mean value and the stochastic fluctuations which
persist from the time when the dissipative processes
$\sigma \leftrightarrow q\overline{q}$ cease.

\section{Simulations of an expanding fireball}
\label{sec:3}

To investigate a situation more close to the medium produced in
heavy-ion collisions we initialize the system as a droplet with a
Woods-Saxon like spherically symmetric temperature distribution
\begin{equation}
\label{3.1}
T(\bvec{x}) = T_0 \left [1+\exp \left(\frac{|x|-R_0}{\alpha} \right)
\right ]^{-1}.
\end{equation}
For this exploratory study we first work with a small system,
corresponding to $R_0=0.45 \; \fm$ and surface thickness $\alpha=0.1 \; \fm$. The
total system size has been chosen to $5 \;\fm$. Also the boundary
conditions have to be adapted from the so far used periodic boundary
conditions to ones that allow for the expansion of the system. For the
particles we choose a ``distance cutoff'' $r_c=2.75 \; \fm$ from the
center, where the particles are removed from the system.

The boundary conditions for the field is more complicated. Ideally one
would choose absorbing boundary conditions (ABC), which perfectly absorb
every wave traveling through the boundary and all its energy  \cite{engquist1977absorbing}.
Unfortunately, such boundary conditions are very hard to implement and
are computational expensive, especially in three dimensions and to have
a good performance their formulation needs to be non-local in time
 \cite{Givoli2004319}. Still, they have been successfully applied for
physical systems like the 3D Schr{\"o}dinger equation
 \cite{PhysRevE.88.053308}.

To keep the computational cost reasonable, the boundary conditions are
therefore kept periodic, but in the outer region of the box the
$\sigma$-field is additionally damped by an effective friction
$\propto \dot \sigma$ for any wave passing the cutoff radius
$| \bvec{x} | > r_c = 2.75 \; \fm$, leading to a strong damping of the
waves. The total energy in this system is not conserved any more and any
energy passing the boundary will be damped or removed from the system,
preventing an interference with the rest of the system.

We have performed simulations for this expanding-fireball situation
using both the model without and with the chemical processes
$\sigma \leftrightarrow \overline{q}q$. The initial temperature was
chosen $T_{\text{initial}}=175 \, \MeV$, the number of test particles
$N_{\text{test}}=10^5$, and a time step of
$\Delta t=2 \cdot 10^{-3} \; \fm/c$.  Neglecting the chemical processes,
the $\sigma$-field shows a rapid oscillation in the beginning of the
expansion and loses energy mainly by radiating long-wavelength
oscillations which propagate out of the system. Without any interaction
processes the system stays nearly isotropic, only small spatial
fluctuations of the quark-density lead to small deviations in the
symmetry.

Additionally, for larger couplings the system can fall into a
meta-stable state in which cold quarks are trapped in a potential well
of the chiral field which has a long lifetime.

Here we concentrate on the results of simulations taking into account
chemical processes $\sigma \leftrightarrow \bar q q$. These processes
should lead to additional particle production whenever the
$\sigma$-field shows strong fluctuations and particle annihilation in
areas of high quark density, inducing additional fluctuations on the
field. There are a couple of qualitative differences between the
calculations with and without chemical processes. For all three
different couplings ($g=3.3$, 3.63, and 5.5 corresponding to a
cross-over, 2$^{\text{nd}}$- and $1^{\text{st}}$-order chiral phase
transition, respectively) the types of fluctuations are very
different. The systems without chemical processes show strong global
fluctuations, leading to shell-like structures with a fast oscillation
of the chiral field at the origin of the matter droplet. The
fluctuations are much less present in the calculations with chemical
processes. The decay process $\sigma \rightarrow \bar q q$ damps such
strong fluctuations by removing kinetic energy from the field. In
contrast, the process $\bar q q \rightarrow \sigma$ creates strong local
fluctuations on the field, increasing with a stronger coupling constant
$g$. Additionally, all calculations without chemical processes stay
nearly isotropic through their time evolution while the calculations
with chemical processes show a breaking of this symmetry after a very
short time. This symmetry is broken by strong and space-dependent
fluctuations of the quark density and the field distribution and due to
a collective drift of both particles and the field disturbance. Such a
drift is surprising but is explained by the conservation of energy and
momenta in the interactions between fields and particles. The
$\sigma$-field can emit particles, generating a momentum-kick in the
opposite direction of the particles, leading to a random-walk phenomena
of the droplet. These outcomes are remarkable because no direct random
processes are involved in the numerical simulation, which would create
these fluctuations. All stochastic and fluctuating processes are the
result of microscopic interaction kernels and cross-sections which are
sampled via Monte-Carlo methods.

\begin{figure*}
  \centering
  \includegraphics[keepaspectratio, width=0.49\textwidth,height=0.25\textheight]{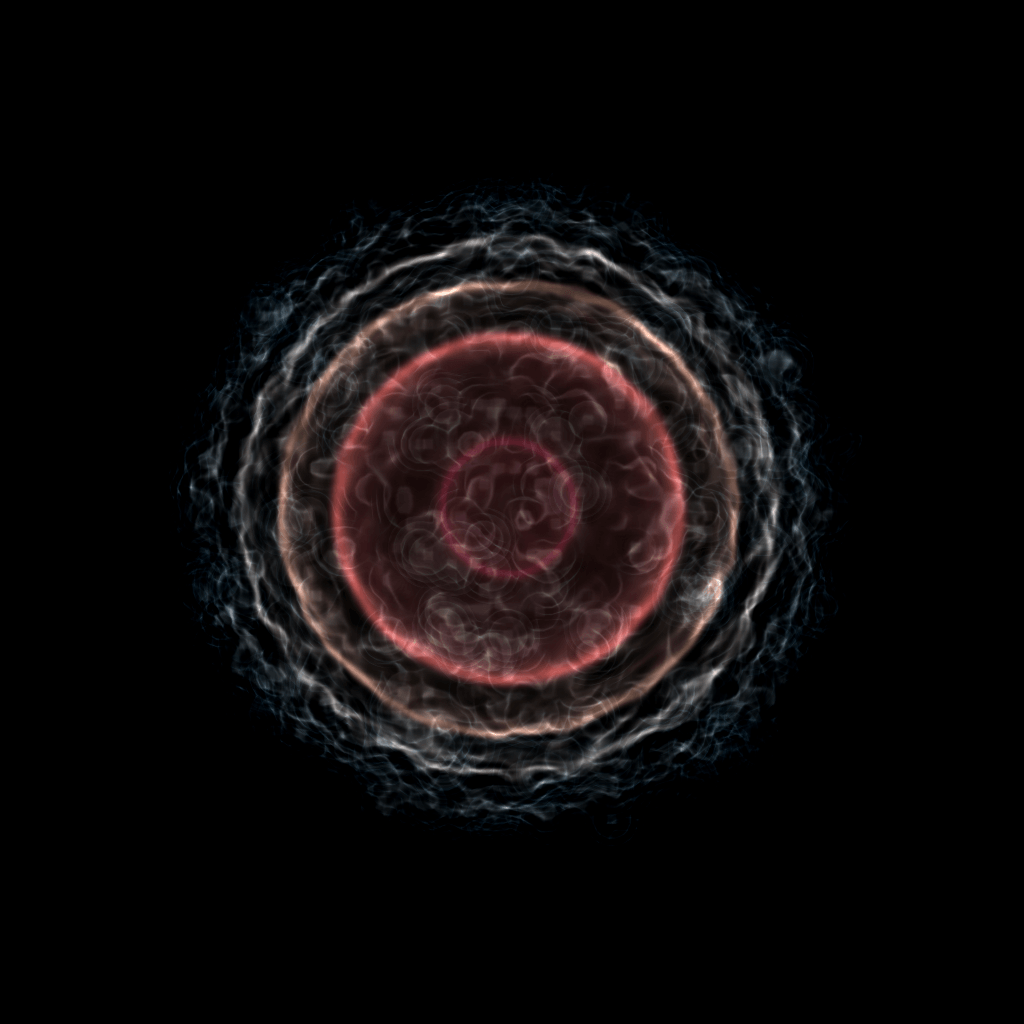}
  \includegraphics[keepaspectratio, width=0.49\textwidth,height=0.25\textheight]{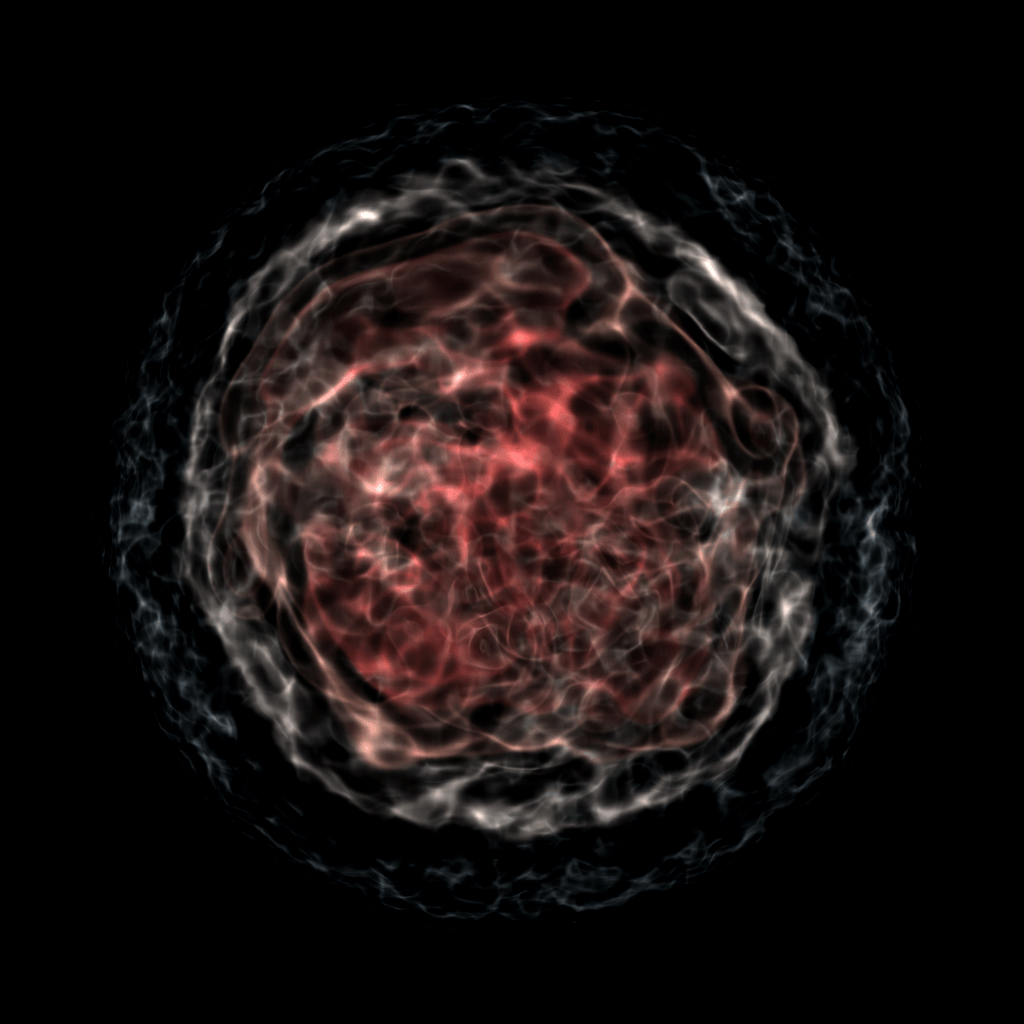}

  \includegraphics[keepaspectratio, width=0.49\textwidth,height=0.25\textheight]{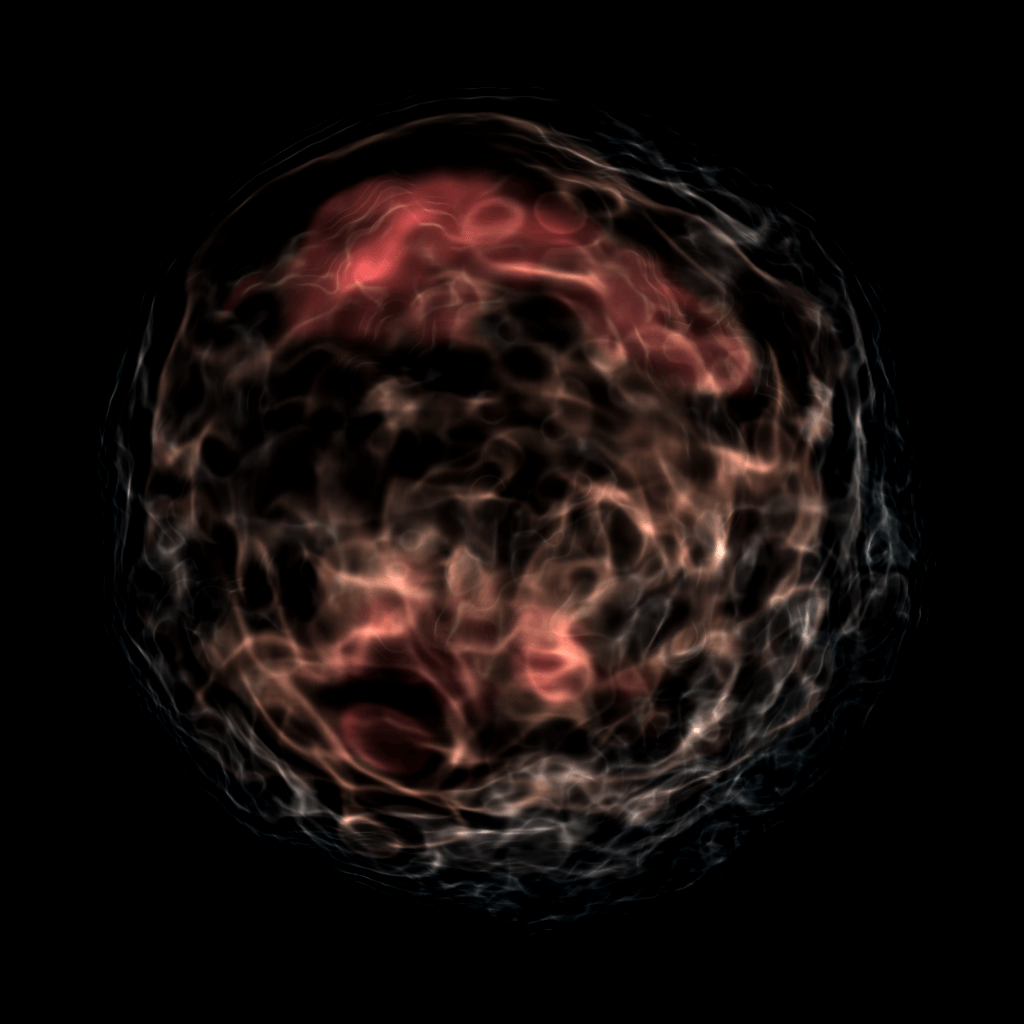}
  \includegraphics[keepaspectratio, width=0.49\textwidth,height=0.25\textheight]{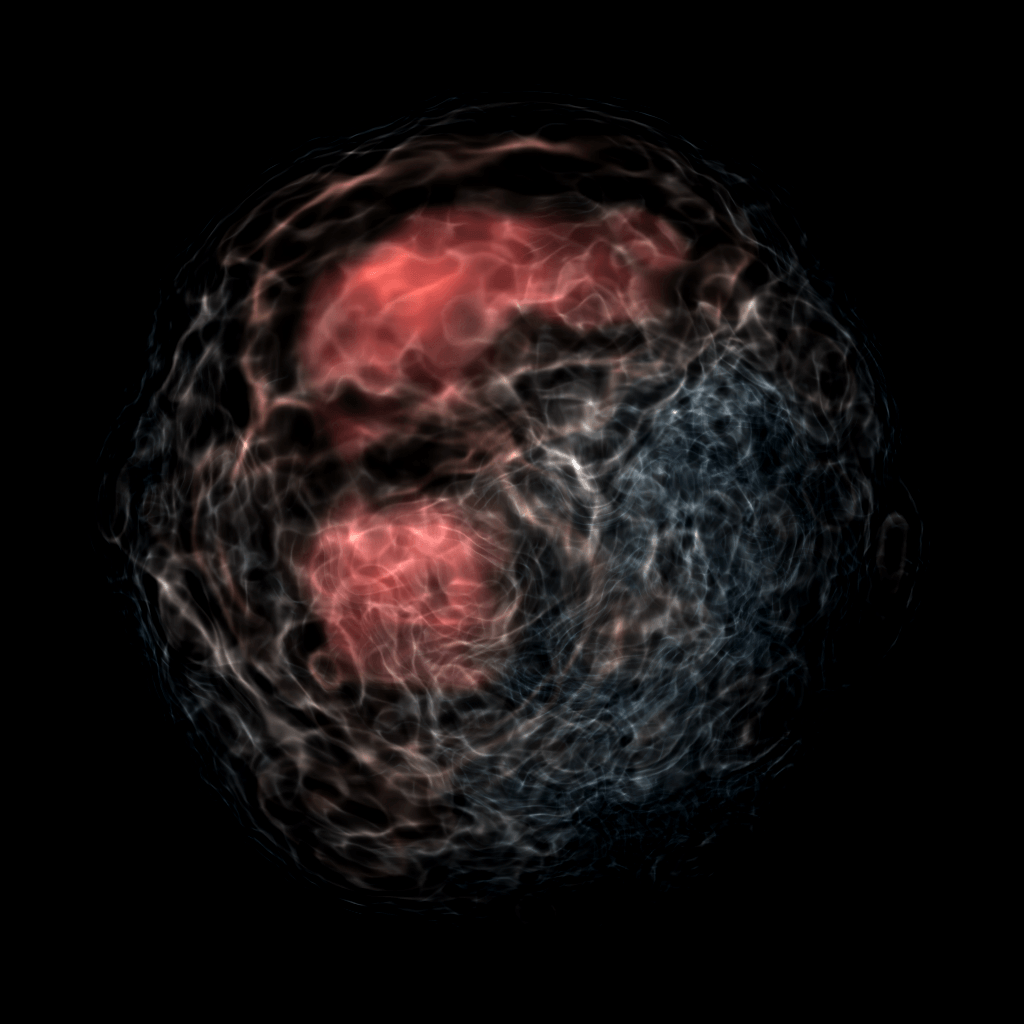}

  \includegraphics[keepaspectratio, width=0.49\textwidth,height=0.25\textheight]{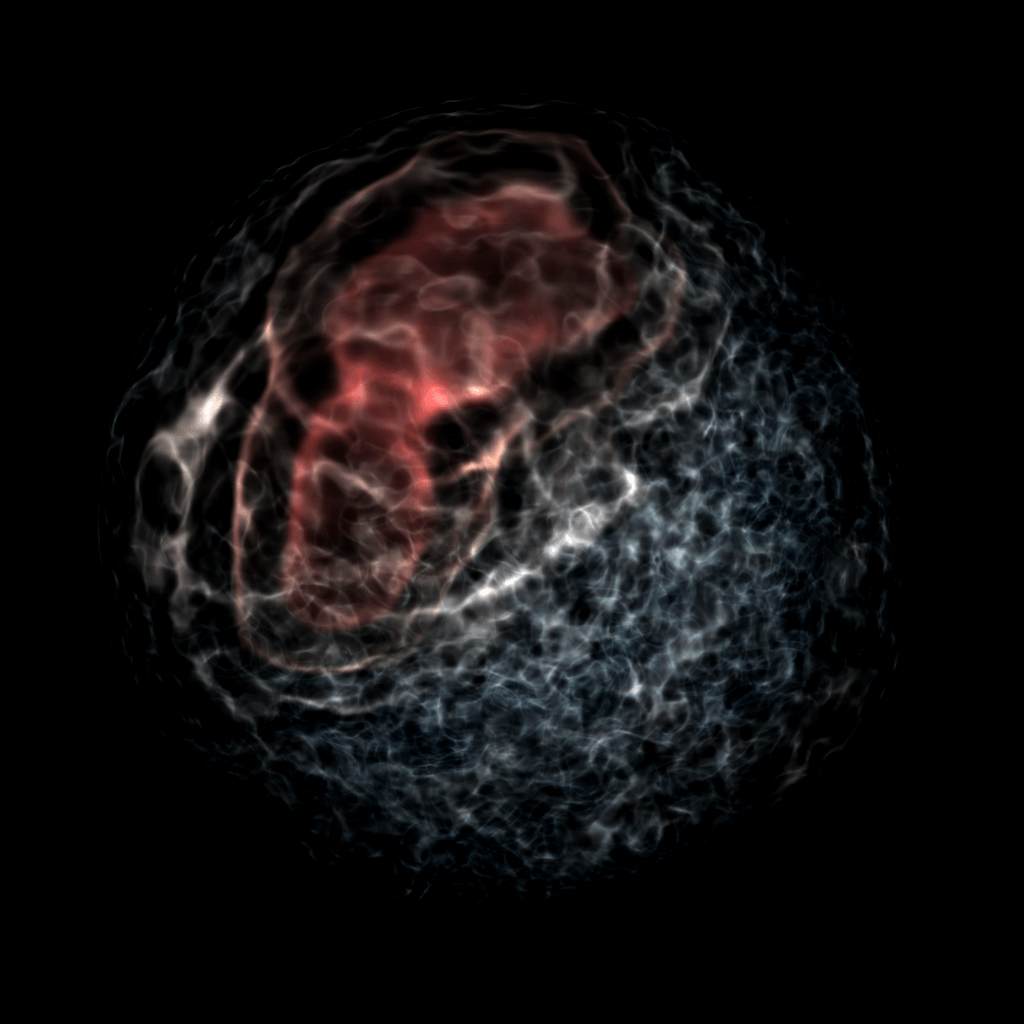}
  \includegraphics[keepaspectratio, width=0.49\textwidth,height=0.25\textheight]{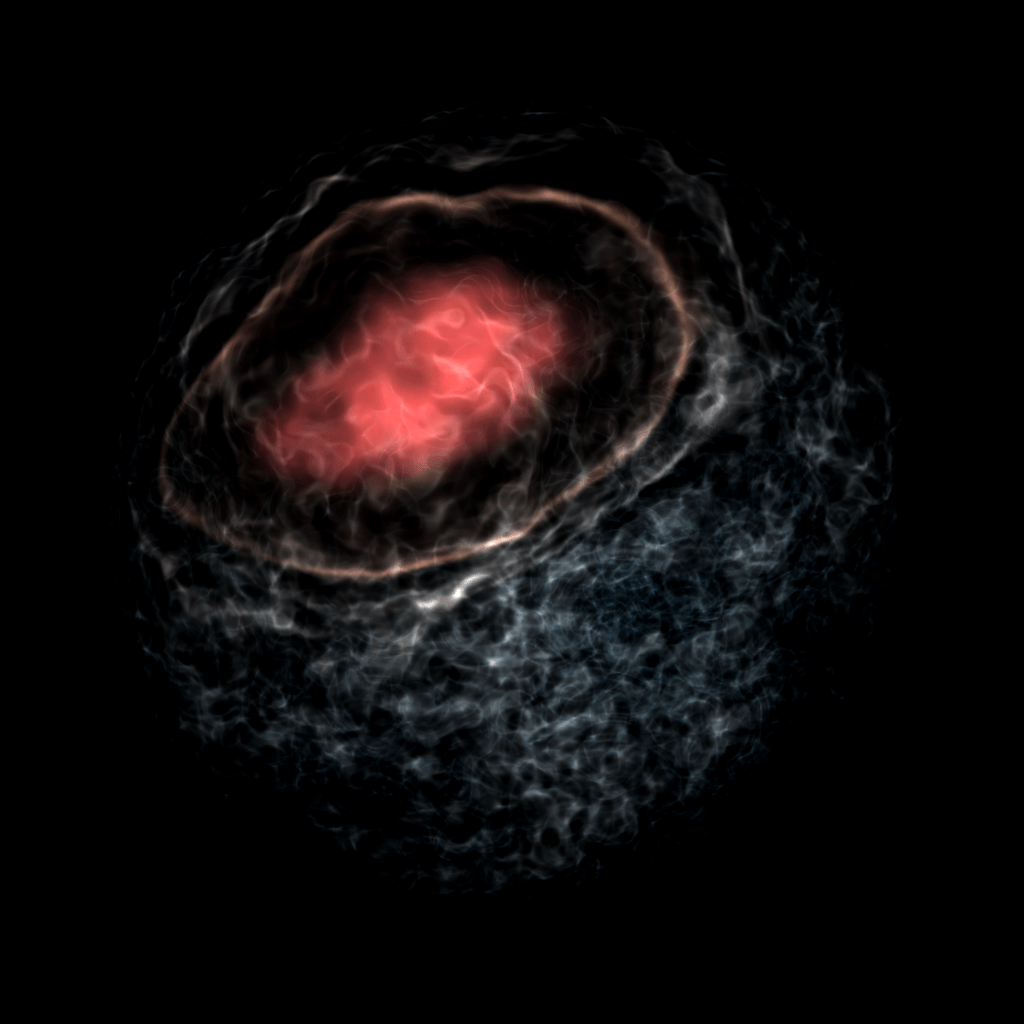}
  \caption{\label{fig.8} Volumetric 3D rendering of the quark density
    for the hot matter droplet scenario with a coupling of $g=5.5$, a
    first-order transition and chemical processes. The time progresses
    for $1 \; \fm / \text{c}$ for every image. Regions with high particle
    densities are colored red, the scale is constant over all
    images. Visualization was created using the yt-toolkit
     \cite{Turk2010}.}
\end{figure*}
By comparing the calculations with chemical processes with different
couplings $g$, qualitative differences are visible, as well. The
fluctuations in the field and particle density increase for stronger
couplings.  For a coupling of $g=5.5$ the quark density forms small
regions with higher densities, which merge with time. These observations
are consistent with calculations of the linear $\sigma$-model with a
hydrodynamic background
\cite{Nahrgang:2011mg,Nahrgang:2011mv,Herold:2013bi}, in which the
authors find the strongest fluctuations for a medium with a first-order
phase transition.  Classical theories of phase transitions in thermal
equilibrium predict the strongest fluctuations at and near the phase
transition for second-order transitions. Calculations presented here
show the strongest fluctuations for the highest coupling associated with
the first-order phase transition. At first, this scenario can not be
directly compared to a phase transition. A phase transition is a
phenomenon described by equilibrium physics for very large systems which
evolve in the adiabatic limit on large time scales. In off-equilibrium
situations fluctuations need time to build up via interactions within
the bulk medium. Most important, the correlation length is often largely
enhanced at the phase transition, which is no problem for systems which
are much larger than this correlation length. The scenario of the
hot-matter droplet is quite the opposite. The system size is in the
order of the interaction length and therefore its correlation
length. The quark matter expands rapidly and its dynamics creates
distributions which strongly deviate from equilibrium, and the total
lifetime of the system is at most in the order of the equilibration
timescale. Additionally, a rapid expansion leads to a highly anisotropic
system with gradients and parts of the system separate to regions with
very different densities and temperatures.  All these circumstances do
not allow a consistent description of the system in terms of an
equilibrium phase transition, especially not if the quarks are described
by particles with non-equilibrium distributions and not by a fast
equilibrating medium. Any deviation from equilibrium descriptions have a
strong impact on the thermal properties of the medium.

Considering the system dynamics in terms of its transport properties
would be a more adequate approach to the characteristics of its
behavior. The strong fluctuations only occur if the chemical processes
$\sigma \leftrightarrow \bar q q$ are present, their microscopic
interaction kernels are derived from a Breit-Wigner
cross-section. Larger couplings lead to a higher interaction rate
between fields and particles (in our model
$\sigma_{\sigma \leftrightarrow \overline{q}{q}} \propto g^2$). This
already implies that the largest fluctuations build up for the
calculations with the coupling $g=5.5$ and quite similar fluctuations
for $g=3.3$ and $g=3.63$, even though these different couplings would
show very different kind of phase transitions in an equilibrated
system. 
\begin{figure*}[t]
\includegraphics[width=0.48\linewidth]{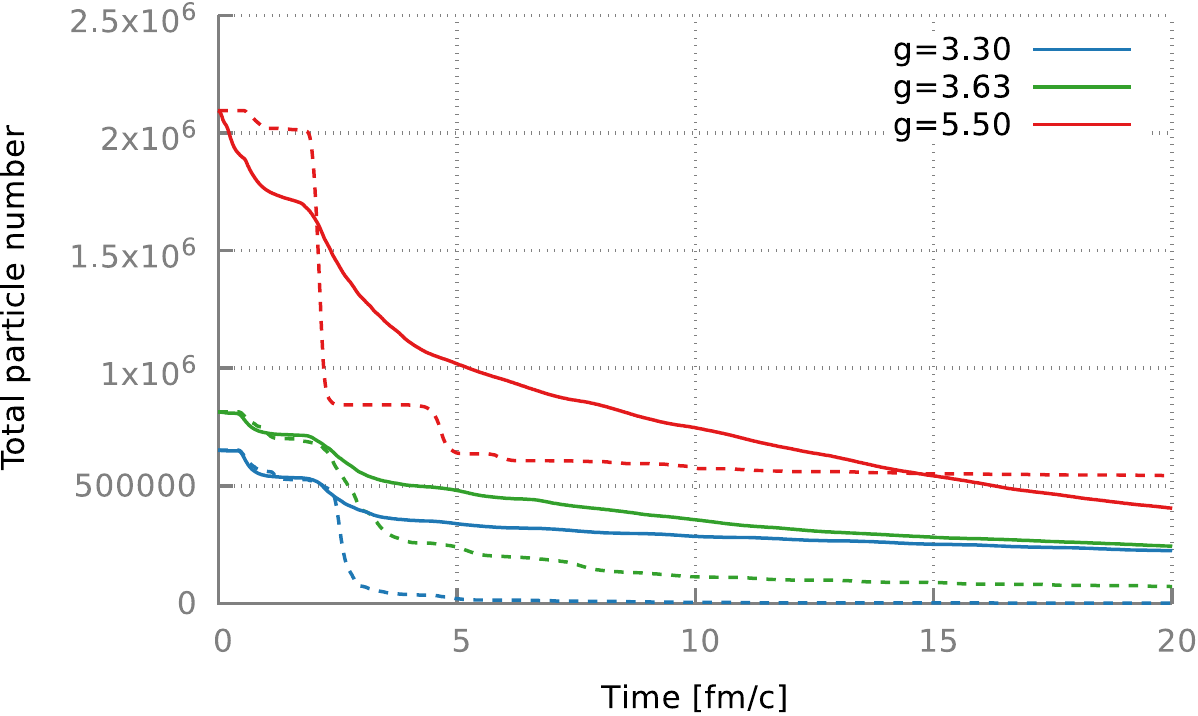}
\includegraphics[width=0.48\linewidth]{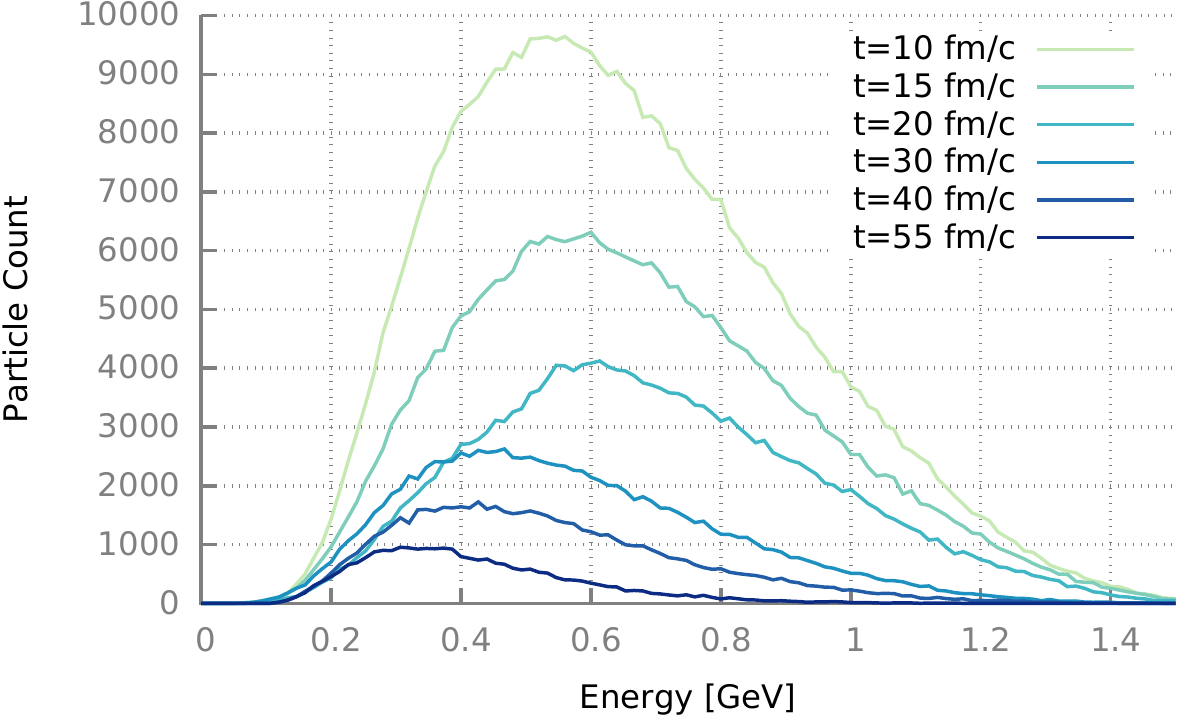}
\caption{(Color online) \textbf{Left:} Total quark number in the
  matter-droplet scenario. Solid lines show the simulations with
  chemical processes, the dashed lines the simulations without the
  $\sigma \leftrightarrow \bar q q$ process. In the scenario without
  chemical processes the droplet radiates most of the quarks in
  shell-like structures, as reflected in the quark-number plateaus which
  drop suddenly. Due to condensation processes the system can form
  meta-stable states, in which cold quarks are trapped in a
  chiral-potential well, which can be observed in the stable, non-zero
  quark number for $g=5.5$ over long times. The behavior is completely
  different for the calculations with chemical processes, in which the
  systems lose quarks in a steady and continuous process, the formation
  of quark-number plateaus is washed out. \textbf{Right:} Energy
  distribution functions of the quarks in the simulation for the
  matter-droplet scenario with $g=5.5$ and chemical processes. After
  $10 \; \fm/c$ most of the initial particles have left the system, and
  the remaining particles have formed a drop of cold quarks, which are
  trapped in a chiral-potential well. One can see a non-thermal
  distribution function in the beginning, which slowly thermalizes due
  to elastic interactions. However, mainly high-energy quarks can leave
  the potential well, leading to both a slow evaporation of the drop and
  to an effective cooling of the remaining particles.}
\label{fig.9}
\end{figure*}

Furthermore, two other aspects play an important role in the comparison
of the different calculations. The coupling has a direct impact on the
mass of the quarks. A higher coupling leads to a larger quark mass in
the chiral broken phase. This implies both a larger particle number at
the same temperature for a higher coupling and a larger potential energy
for the chiral fields. The second aspect has its origin in the phase
diagram. Higher couplings lead to a lower $T_c$ in the linear
$\sigma$-model, which strongly changes the dynamics of the chemical
interactions as they are only possible above the mass threshold, meaning
above $T_c$. For lower $T_c$, in comparison to other couplings, the
quarks and the fields have more time to stay in the chiral restored
phase, more time to interact with each other and therefore more time to
build up fluctuations via these interactions. This already implies
stronger fluctuations, regardless of the type of phase transition, which
would be given by the corresponding coupling.  A fair comparison between
the scenarios is not given by comparing the system at different
temperatures. Better approaches could be a comparison with same energies
or same particle number.

In the calculations neglecting chemical processes the formation of
meta-stable drops of quarks could be observed, especially at the high
coupling, $g=5.5$. This behavior can be seen in the simulations
including chemical reactions, too. For $g=5.5$ with chemical reactions
the system even shows something like bubble formation.  Figure
\ref{fig.8} shows a volumetric 3D rendering of the quark density which
projects the full three dimensional density and visualizes how small
areas of high quark-density start to merge into larger areas, having
some similarity to the condensation of water drops in steam. The
mechanism of this condensation is that it is the energetically favorable
configuration for the $\sigma$-field. Interestingly, the condensation
progresses and the $\sigma$-field looses energy by radiating wave
excitations. The process seems to stop after $10 \; \fm/\text{c}$ by
forming a quasi-stable drop. The left panel in Figure \ref{fig.9} shows
the total quark number of the system, showing a steady decrease of
quarks in the system, indicating a kind of particle evaporation from the
drop. This evaporation seems to be quite slow and calculations have
shown a lifetime of up to $50 \; \fm/\text{c}$ before the drop
bursts. The right panel Figure \ref{fig.9} displays the evolution of the
quark-distribution function with time after a condensation drop has been
formed. The figure shows a decrease of the quark number with time, and
mainly particles with high energies leave the system. This is reasonable
because particles with high energy can escape the potential well,
leading to a collective cooling of the remaining medium.

All calculations discussed so far are done for systems with a fireball
which has an initial diameter of roughly $d \approx 1 \; \fm$. In
comparison to a heavy-ion collision such a system is very small. Both
lead and gold nuclei have a diameter of around
$d_\text{Au} \approx 14 \; \fm$, resulting in a volume two orders of
magnitude larger than in this scenario. Therefore quantified predictions
for a heavy-ion collision can not be derived from these calculations.
Both the number of involved quarks and the energy density in the system
could lead to different phenomena. Additionally, a much larger spatial
and time scale may change the behavior at the phase transition because
the system has potentially more time to evolve and form structures in
the particle densities.

Now we simulate the model with a larger system volume of
$V=(36 \; \fm)^3$ with an initial size of the matter droplet of
$d \simeq 14 \; \fm$, roughly resembling a central Au-Au collision. The
initial temperature was again chosen to be $T=175 \; \MeV$, as well as
the couplings $g=3.3$, $g=3.63$, and $g=5.5$ to investigate the cases of
cross-over, 2$^{\text{nd}}$, and $1^{\text{st}}$ order phase
transitions. The grid size is increased to $N_{\text{grid}}=192^3$, and
the time step is reduced to $\Delta t=10^{-3} \; \fm/c$. Also the
interaction volume has been slightly increased to
$V_{\text{interaction}}=(0.4 \; \fm)^3$. The total number of particles
was $\sim 5 \cdot 10^6$ with a test-particle multiplicator of $200$.

\begin{figure*}
   \centering
   \begin{subfigure}[t]{0.4\textwidth}
       \includegraphics[width=\textwidth,keepaspectratio]{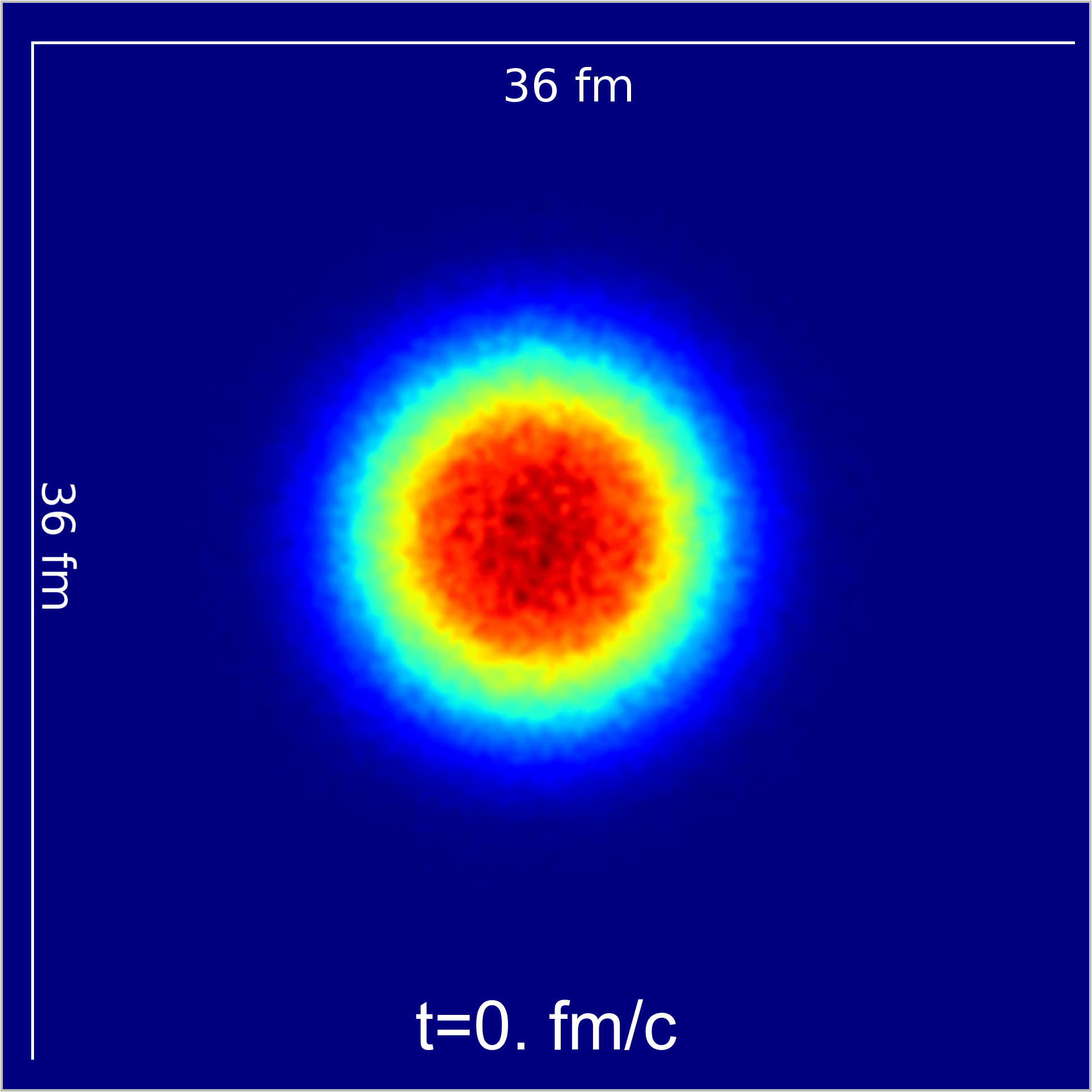}
       \includegraphics[width=\textwidth,keepaspectratio]{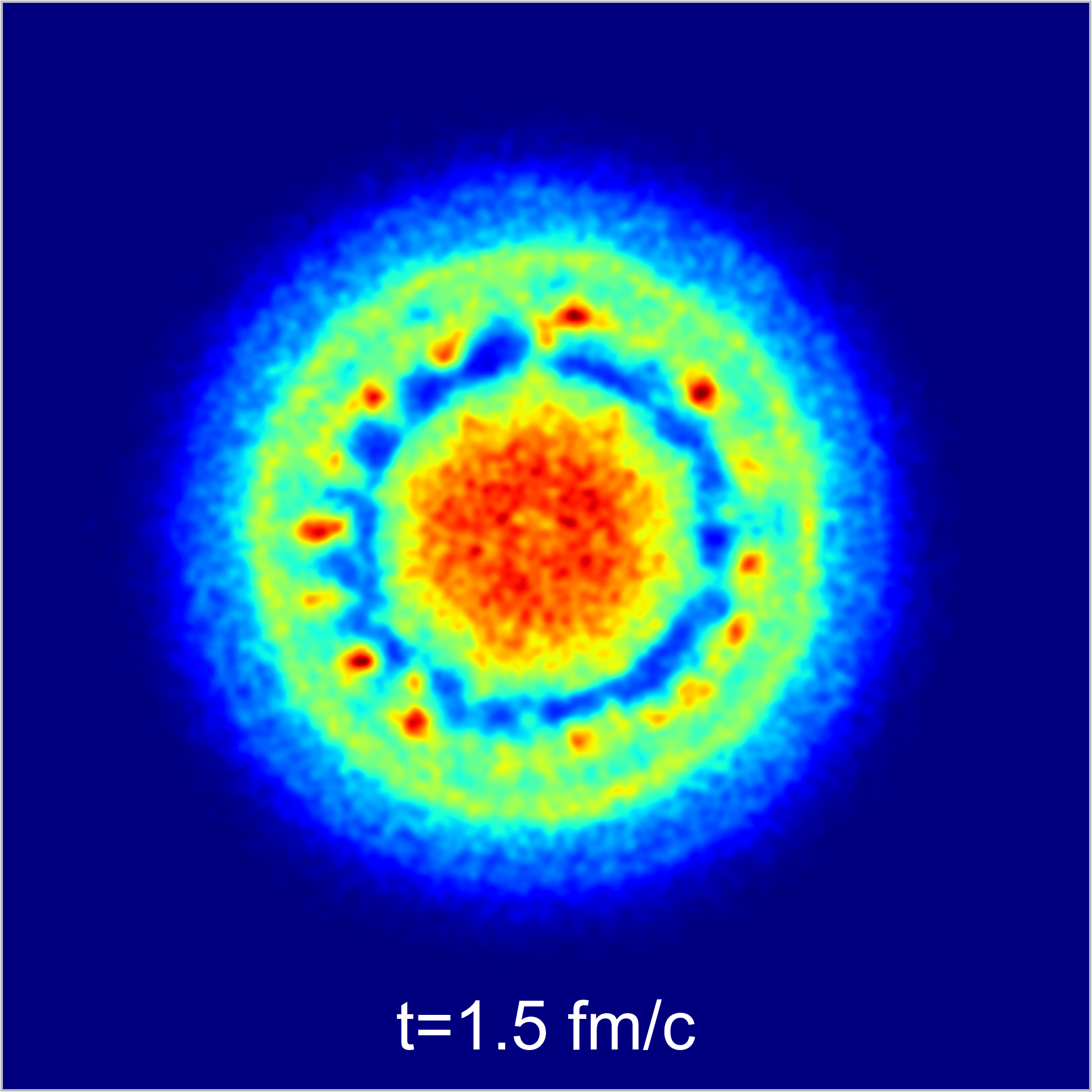}
       \includegraphics[width=\textwidth,keepaspectratio]{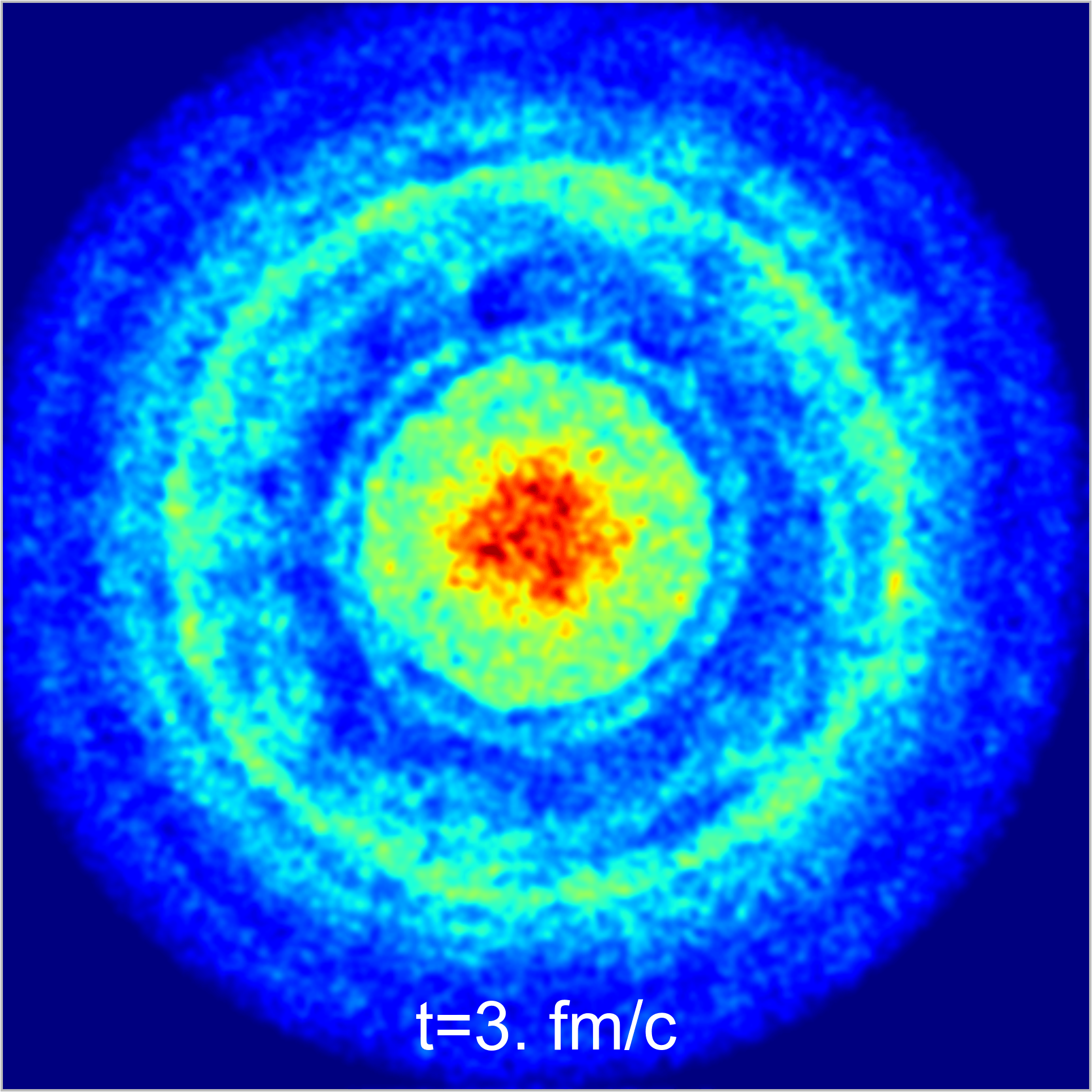}
   \end{subfigure}
    \hfill
   \begin{subfigure}[t]{0.4\textwidth}
       \includegraphics[width=\textwidth,keepaspectratio]{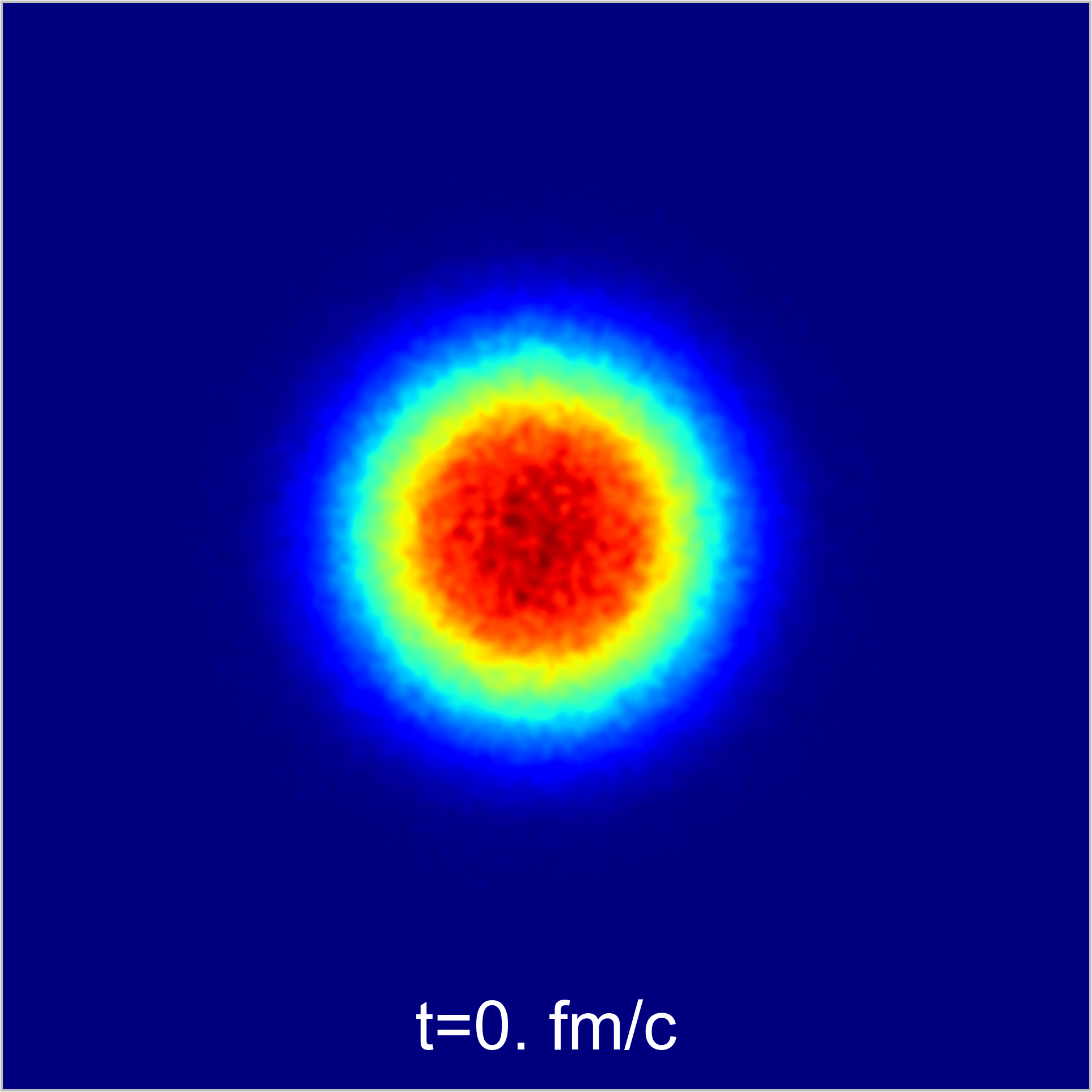}
       \includegraphics[width=\textwidth,keepaspectratio]{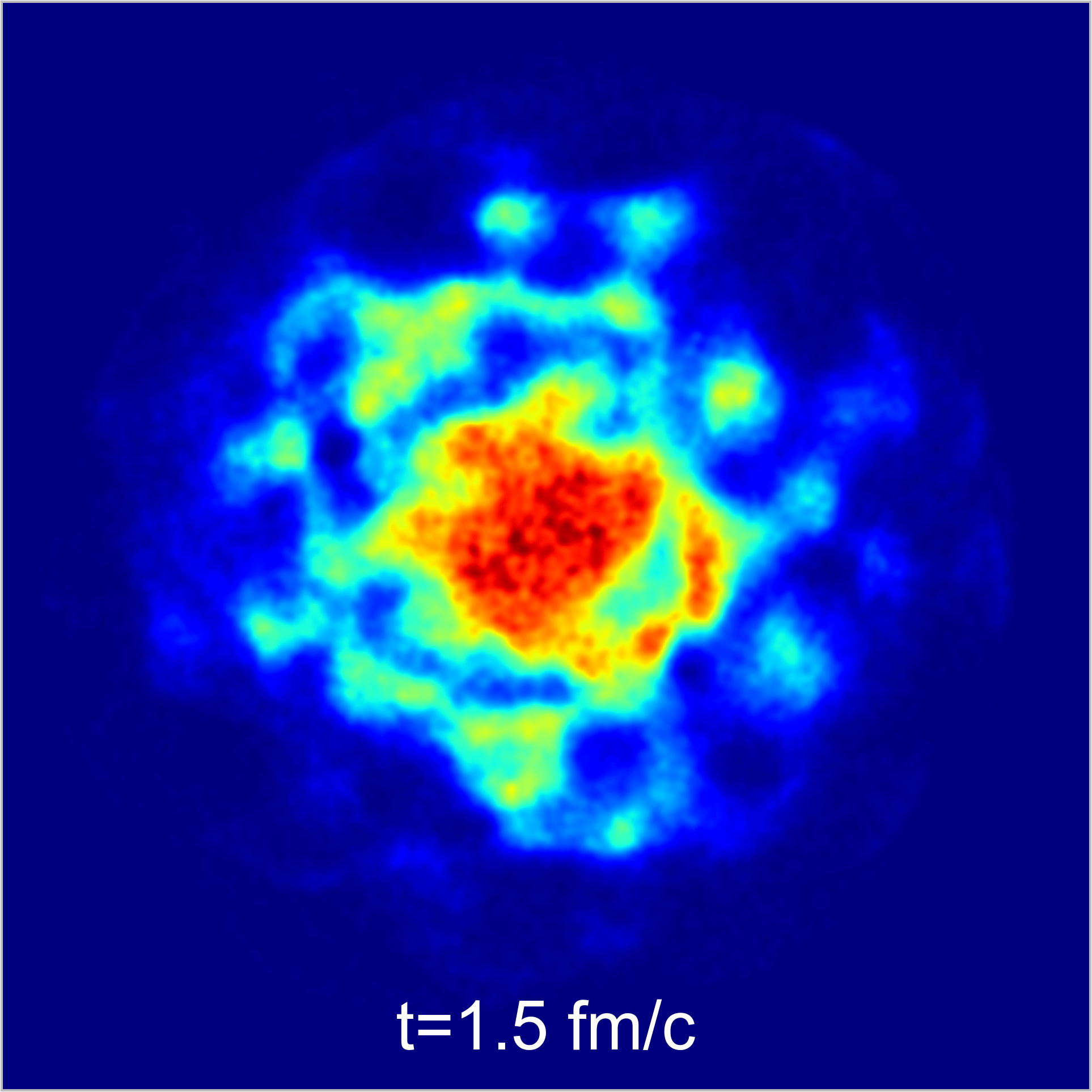}
       \includegraphics[width=\textwidth,keepaspectratio]{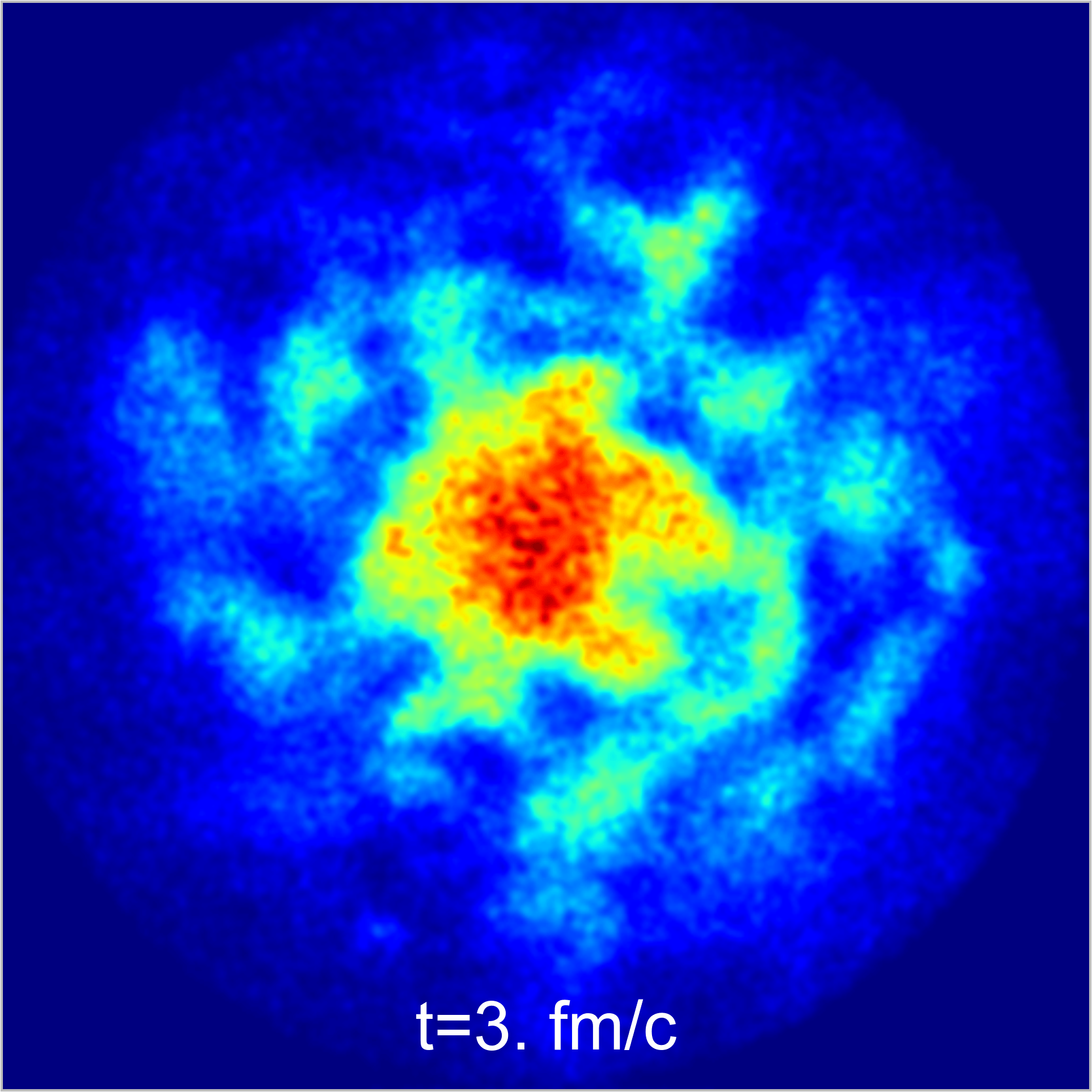}
   \end{subfigure}
   \caption{\label{fig:calcs:physical:largeBlobs} Evolution of the quark
     density for a large hot matter-droplet scenario with $g=3.3$. The
     initial size of the droplet has an diameter of around
     $d \approx 14 \; \fm$ which corresponds to the size of a gold
     nuclei, the total size of the system is $V = (36 \;
     \fm)^3$. \textbf{Left:} Simulation without chemical interactions
     \textbf{Right:} simulation with chemical interactions. Chemical
     interactions lead to the formation of anisotropic structures, but
     the overall expansion of the system is not strongly altered.}
\end{figure*}

\begin{figure*}
   \centering
   \begin{subfigure}[t]{0.4\textwidth}
       \includegraphics[width=\textwidth,keepaspectratio]{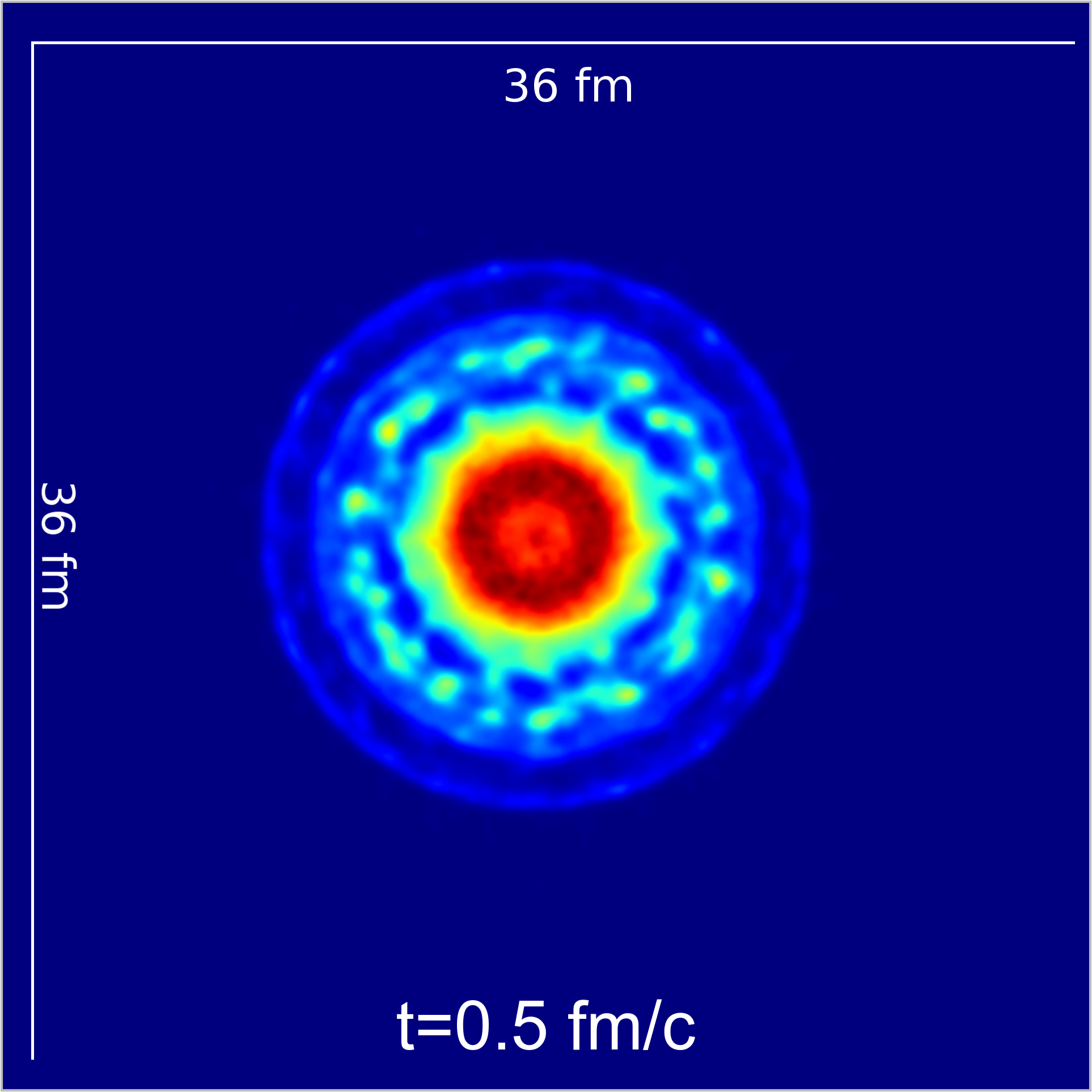}
       \includegraphics[width=\textwidth,keepaspectratio]{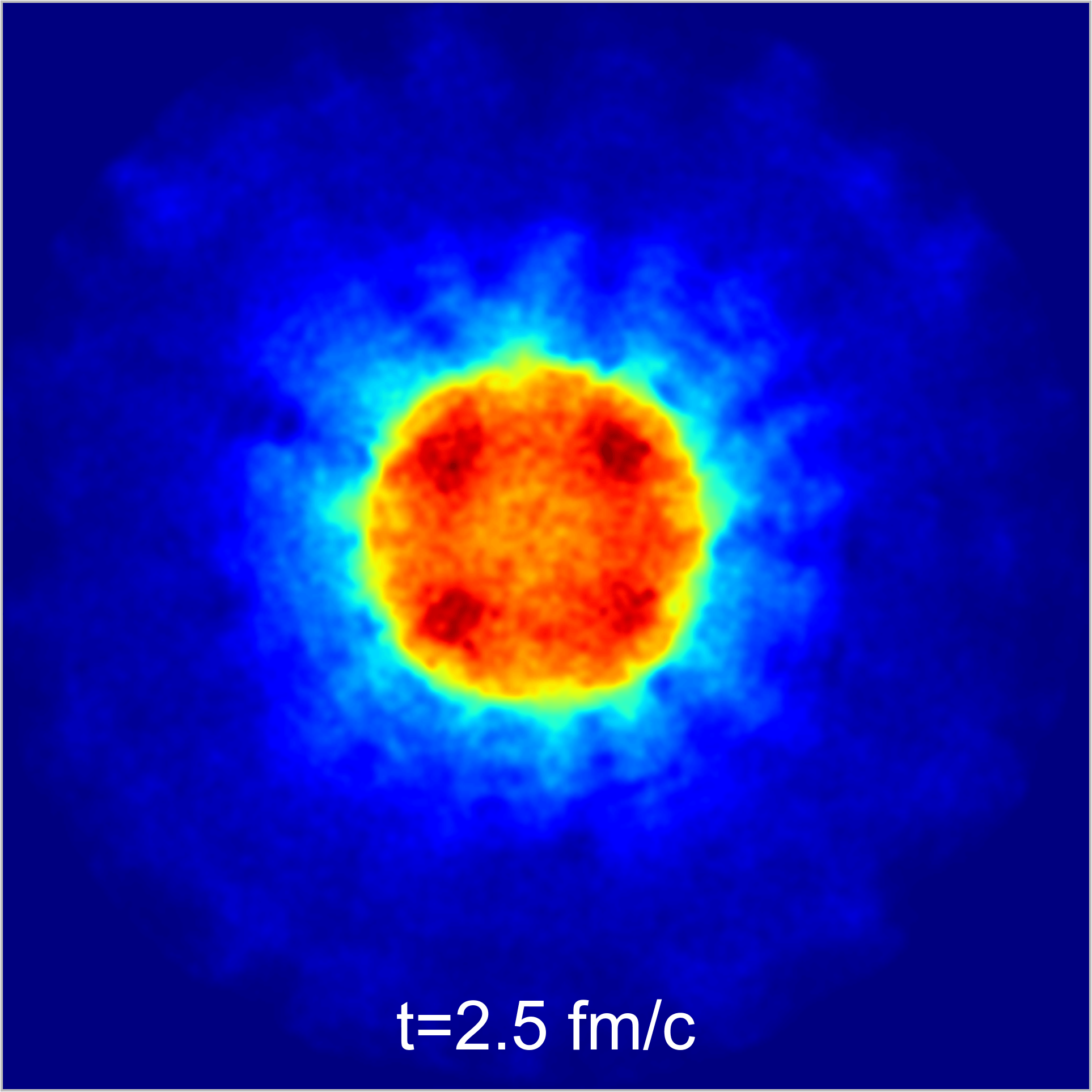}
       \includegraphics[width=\textwidth,keepaspectratio]{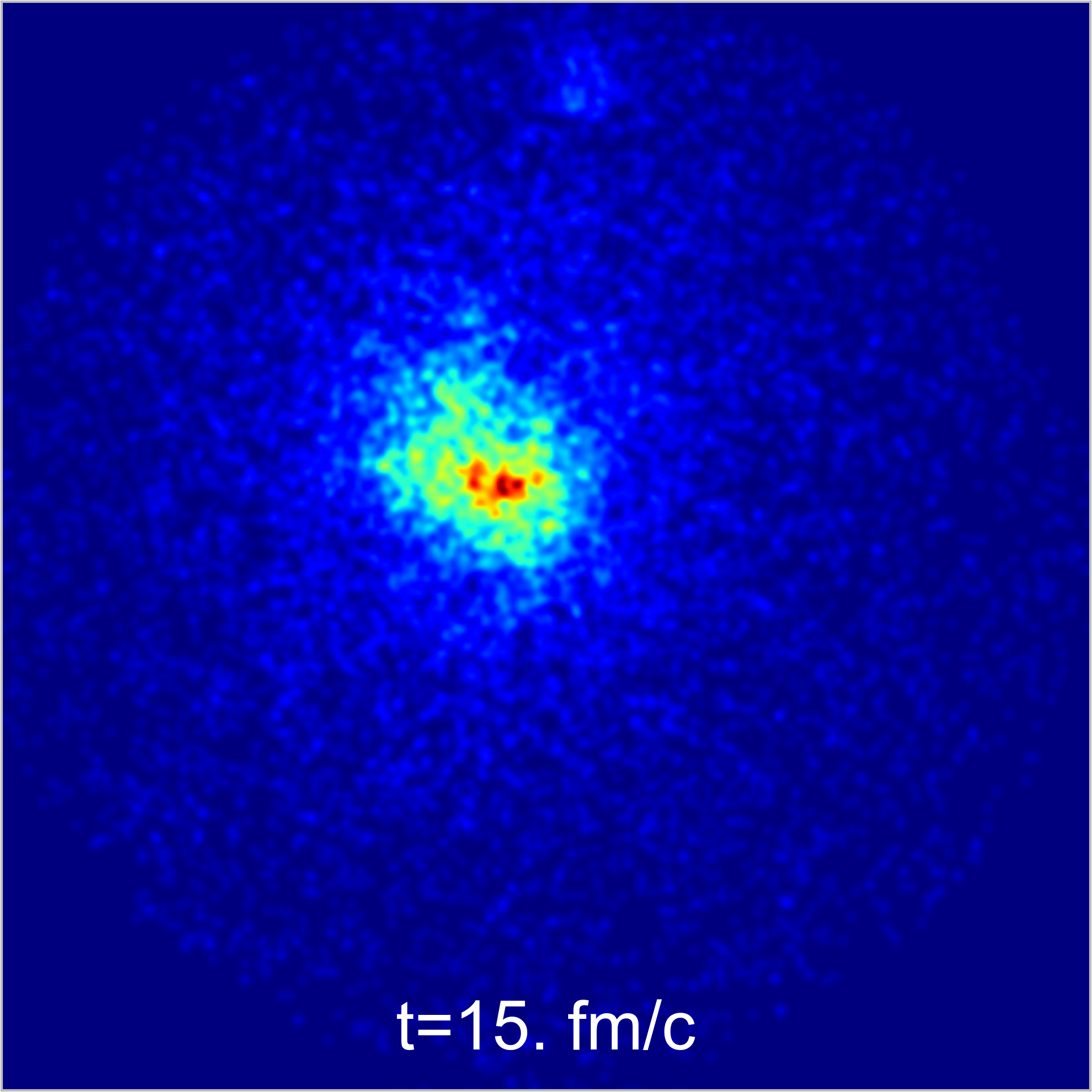}
   \end{subfigure}
    \hfill
   \begin{subfigure}[t]{0.4\textwidth}
       \includegraphics[width=\textwidth,keepaspectratio]{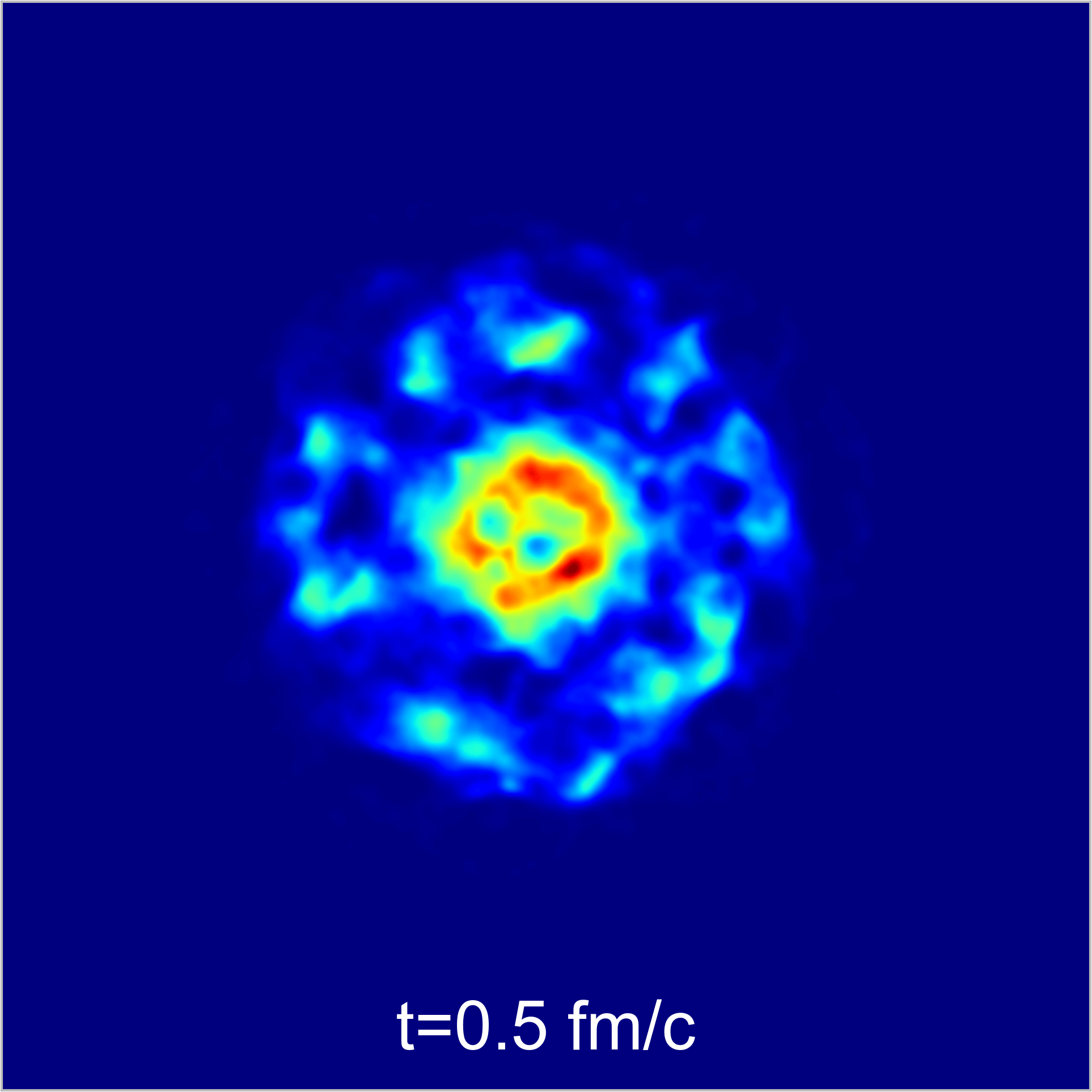}
       \includegraphics[width=\textwidth,keepaspectratio]{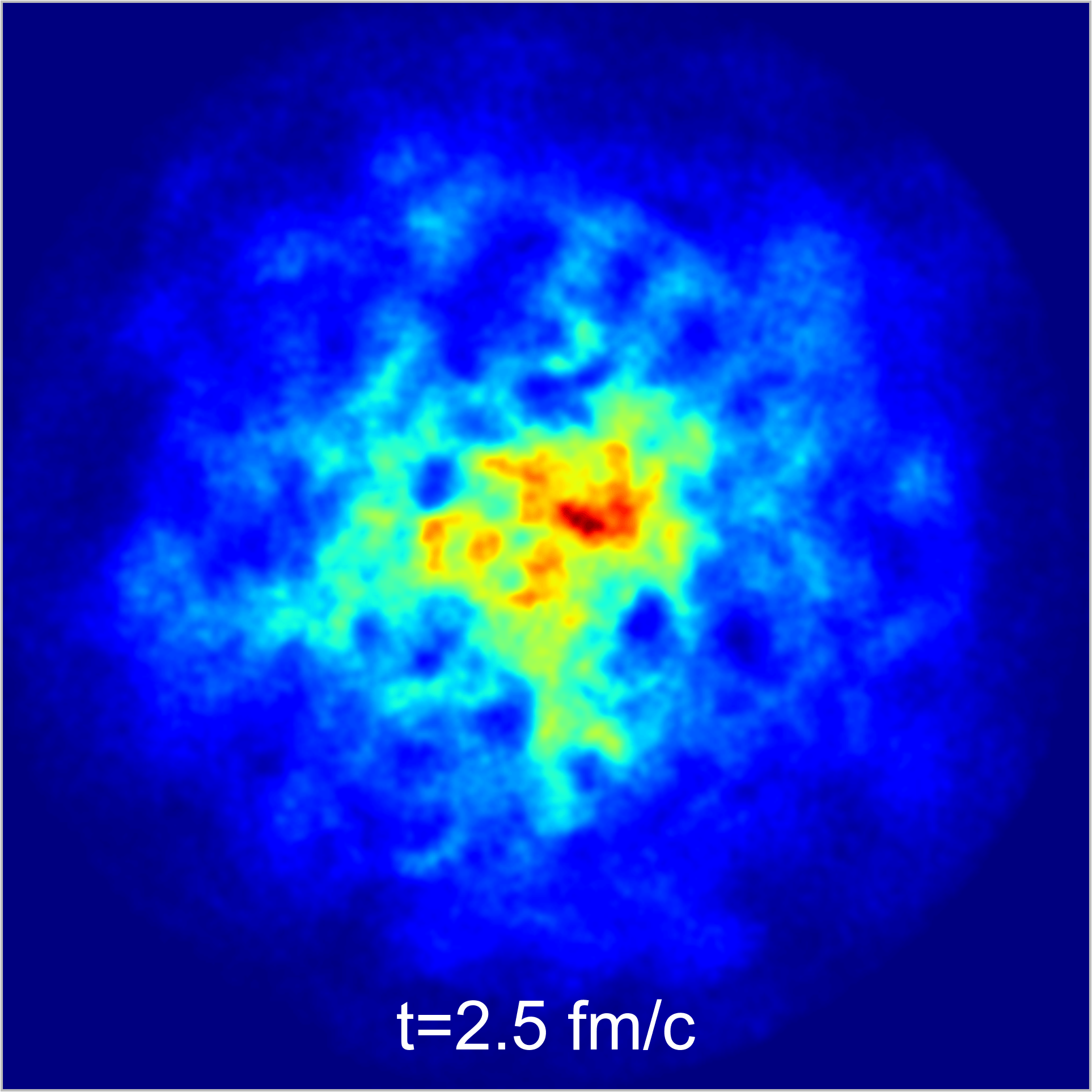}
       \includegraphics[width=\textwidth,keepaspectratio]{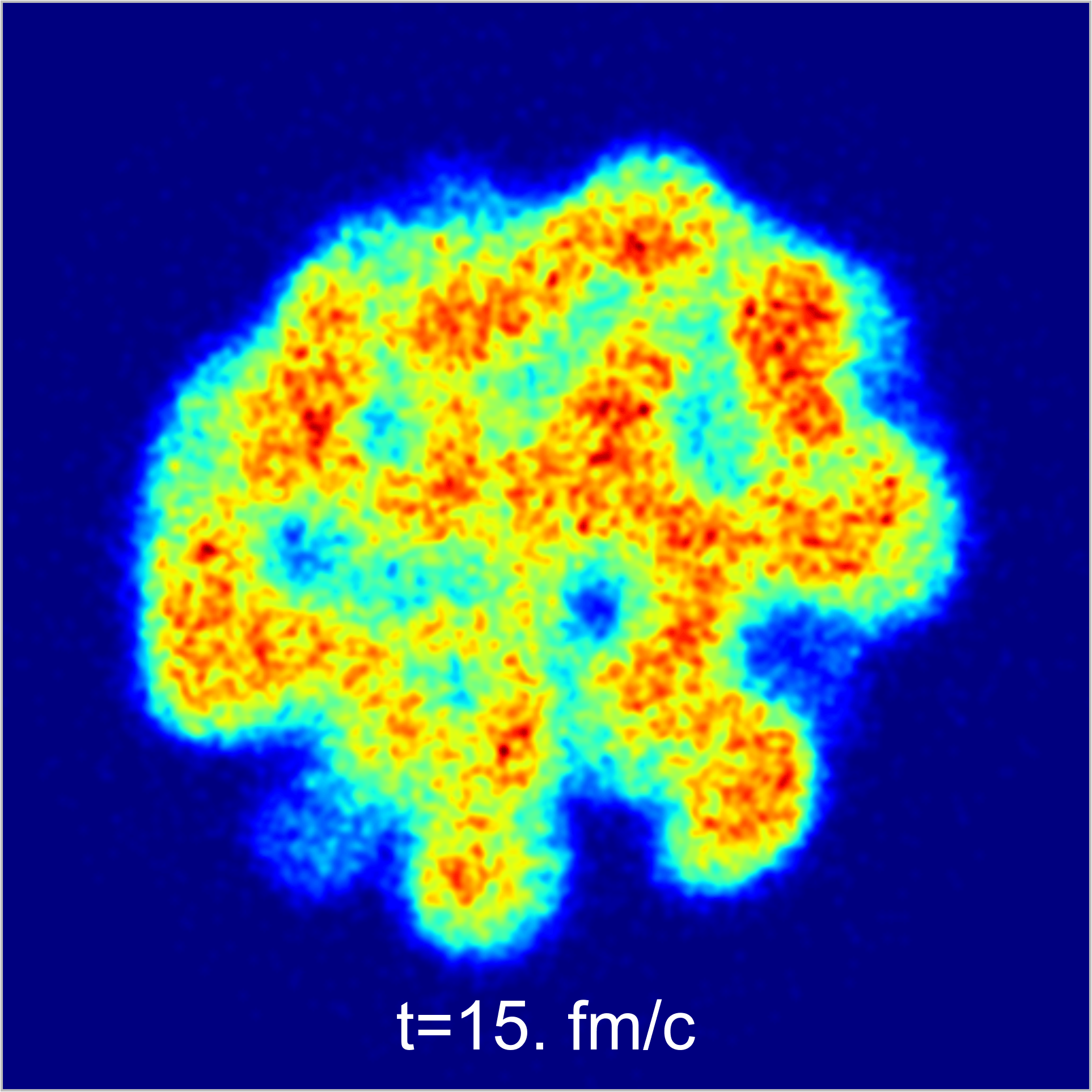}
   \end{subfigure}
   \caption{\label{fig:calcs:physical:largeBlobs2} Quark density for a
     large hot matter-droplet scenario with $g=5.5$. The initial size of
     the droplet has a diameter of about $d \approx 14 \; \fm$ which
     corresponds to the size of a gold nucleus, the total size of the
     system is $V = (36 \; \fm)^3$.  \textbf{Left:} Simulation without
     chemical interactions \textbf{Right:} simulation with chemical
     interactions. Chemical interactions lead to a generation of strong
     and detailed structures. After $5 \; \fm/c$ the expansion of the
     system is stopped by the strong damping of the field, leading to
     the formation of local bubbles.}
\end{figure*}

Figure \ref{fig:calcs:physical:largeBlobs} and
\ref{fig:calcs:physical:largeBlobs2} show the result for a calculation
with a large expanding matter droplet. Both figures compare the same
calculations with and without chemical interactions and different
couplings, $g=3.3$ with a cross-over transition in Figure
\ref{fig:calcs:physical:largeBlobs} and $g=5.5$ in Figure
\ref{fig:calcs:physical:largeBlobs2}.

The calculations without chemical interactions do not change very much
in comparison to the smaller systems. The overall symmetry of the system
stays effectively intact, and known structures like the shell-structures
\cite{Abada:1994mf,Greiner:1996md} and cold quark droplets are still
present.

The qualitative picture changes somewhat for calculations with chemical
interactions, especially for larger couplings like $g=5.5$. After a
short expansion phase the matter droplet starts to form strong and
non-isotropic structures in the quark density due to the annihilation
and creation process. In Figure \ref{fig:calcs:physical:largeBlobs}
first structures are already present at $t \approx 1 \,
\text{fm}/c$. These structures blur while the matter expands.

In Figure \ref{fig:calcs:physical:largeBlobs2} these structures are much
stronger and much finer. The reason is the strong coupling between the
field and particles, leading to fast annihilation and strong damping of
the field by $\sigma$-decay. While expanding, the fluctuations on the
$\sigma$-field become stronger. Around $t=5 \, \text{fm}/c$ the
annihilation-rate of the quarks is negligible because of the low
particle density due to expansion. The $\sigma$-decay becomes the
dominating effect, leading to a strong damping of the field and to a
quasi-freeze out of the field leading to some kind of local
bubble-formation. The resulting field distribution stays stable for
several $10 \; \fm/c$ local bubbles converge slowly to a big
drop. This final drop contains the already known cold quarks which were
not able to escape the kinetic potential.

In the calculations for larger systems a qualitative difference between
the couplings can be seen in the time evolution. While calculations for
$g=3.3$ and $g=3.63$ behave similar, calculations for $g=5.5$ show a
very different behavior: the quark matter forms bubbles and freezes out
in a relative long-living structure.

\begin{figure*}[t]
\centering
\begin{minipage}{0.48\linewidth}
\includegraphics[width=0.99\linewidth]{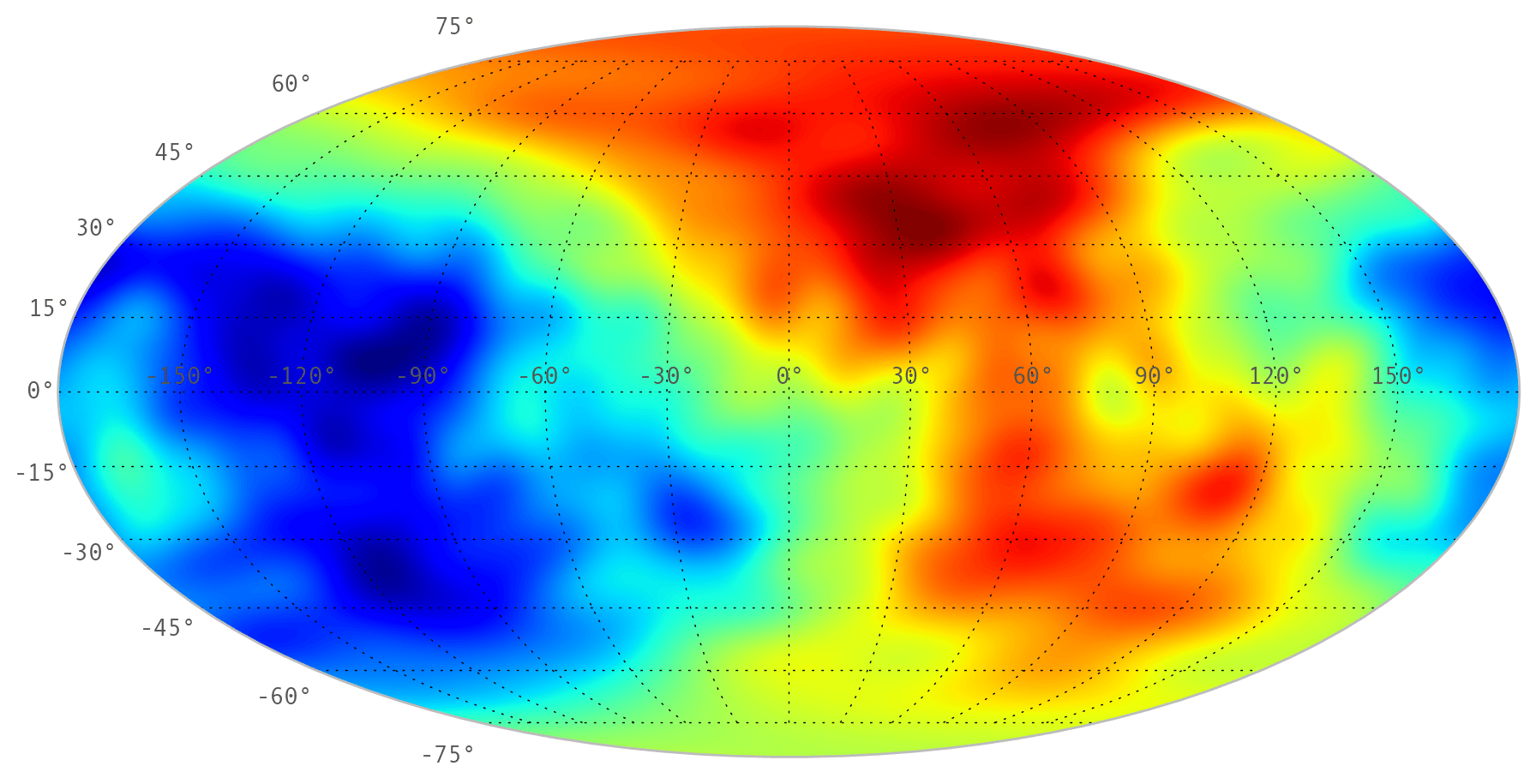}
\end{minipage}
\begin{minipage}{0.48\linewidth}
\includegraphics[width=0.99\linewidth]{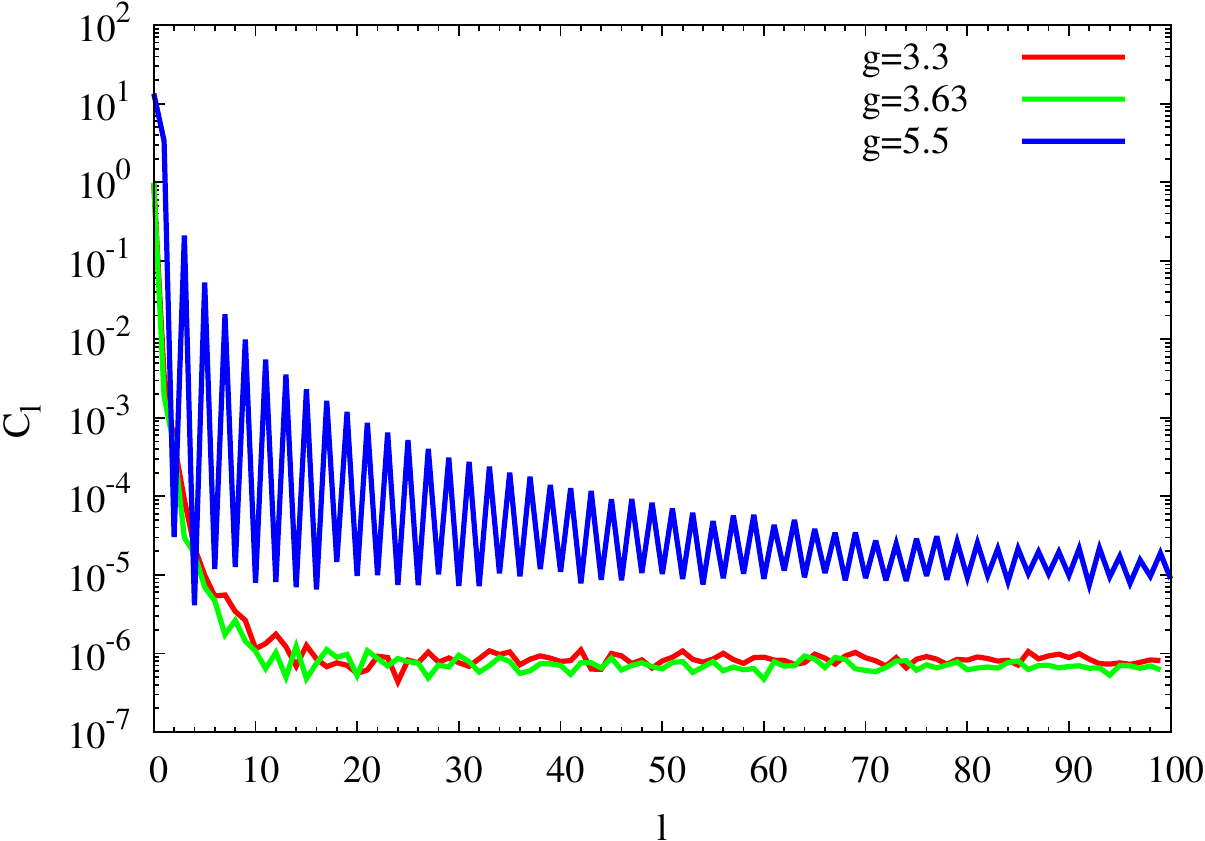}
\end{minipage}
\caption{(Color online) Angular distribution (\textbf{left panel}) of
  quarks leaving the fireball for the coupling $g=3.3$ and the power
  spectrum of the angular correlation functions (\textbf{right panel}).}
\label{fig.10}
\end{figure*}

The calculations in this section have shown the very interesting
complexity of such a simple initial condition like an expanding matter
droplet. The chemical reactions between particles have a very strong
impact on the system behavior and dramatically change both fluctuations
and medium propagation within the expansion.  However, characteristic
signatures which would allow an event-by-event discrimination between
the different coupling strengths (implying different phase transitions
in equilibrium) in this scenario has not been found, at least not for
calculations with the cross-over coupling $g=3.3$ and the second-order
transition coupling $g=3.63$. As a first attempt to quantify the
quark-number fluctuations in Fig.\ \ref{fig.10} we plot the angular
momentum distribution of quarks leaving the fireball volume for the case
of $g=3.3$. The qualitative features of the distribution do not
significantly vary with the coupling strength. Only for the first-order
value $g=5.5$ show significantly larger magnitude of the fluctuations,
as quantified by the power spectra of the angular correlation function.

To quantify the angular distributions further, we evaluate the angular
power spectrum, via
\begin{equation}
\label{ang-cor.1}
C_{lm} = \int_{\Omega} \dd \Omega f(\vartheta,\varphi) \text{Y}_{lm}(\vartheta,\varphi)
\end{equation}
with the usual normalized spherical harmonics
$\text{Y}_{lm}(\vartheta,\varphi)$ and 
\begin{equation}
\label{ang-cor.2}
f(\vartheta,\varphi)=\int_0^{\infty} \dd p \frac{\dd}{\dd^2 p} N_{\text{quark}}(p,\vartheta,\varphi).
\end{equation}
The normalized power spectrum is then defined by
\begin{equation}
\label{ang-cor.3}
C_l=\frac{1}{2l+1} \sum_{m=-l}^l |C_{lm}|^2.
\end{equation}

\section{Conclusions and outlook}
\label{sec:4}

In this paper we have used a novel numerical algorithm to simulate the
off-equilibrium dynamics of a quark-meson linear-$\sigma$ model (DSLAM)
inspired by ``wave-particle duality'' of ``old quantum mechanics'',
guaranteeing accurate energy-momentum conservation and the principle of
detailed balance by treating the quarks and anti-quarks, coupled to a
$\sigma$ mean-field with the test-particle Monte-Carlo approach to
evaluate the two-body elastic-collision terms and a combined method
using wave-particle duality and coarse graining to evaluate chemical
processes of quark-antiquark pair annihilation and $\sigma$-meson decay,
$q \overline{q} \leftrightarrow \sigma$. On top also the mean-field
equation for the $\sigma$ field has been solved consistently.

The algorithm has been validated in ``box calculations'' by checking the
energy-momentum conservation, the stability of the expected equilibrium
solutions for the mean field and quark-antiquark phase-space
distribution functions, and the approach to the correct thermal and
chemical equilibrium in the long-time limit starting with simple
off-equilibrium initial conditions (``thermal quench''), demonstrating
excellent fulfillment of energy-momentum conservation and of the
principle of detailed balance. The implementation of the chemical
processes $q \overline{q} \leftrightarrow \sigma$ are found to be
crucial for the correct description of the phase transition and the
approach of the equilibrium state as expected from the corresponding
equilibrium quantum-field theoretical evaluation of the model on the
mean-field level.

Afterward the model has been applied to the simulation of expanding
fireballs mimicking the behavior of the strongly interacting medium as
created in heavy-ion collisions. We have found that with this model we
can simulate the expected local quark-antiquark, i.e., baryon number
fluctuations (under strict obedience of the global baryon-number
conservation law) due to the implementation of the quark-antiquark
annihilation and $\sigma$-meson decay processes. Particularly for
strong couplings of the $\sigma$-meson to the quarks and antiquarks,
corresponding to a $1^{\text{st}}$-order phase transition in
equilibrium, quasi-stable bubbles of ``cold'' quarks and anti-quarks,
trapped in the chiral potential due to the mean field has been
observed. As a first attempt to find a possible observable in heavy-ion
collisions for these fluctuations we have evaluated the angular distributions
of the quarks and antiquarks as a putative measure for net-baryon-number
fluctuations and the associate power spectrum of the angular-density
correlations. Unfortunately we have not found qualitative differences in
the shape of these power spectra for different couplings associated with
cross-over, $1^{\text{st}}$, and $2^{\text{nd}}$ order chiral phase
transitions but only in the absolute strength of the fluctuations as
expected from the associated different coupling strengths between
$\sigma$ mesons and quarks/antiquarks.

\textbf{Acknowledgment:} This work was partially supported by the
Bundesministerium f{\"u}r Bildung und Forschung \linebreak (BMBF
F{\"o}rderkennzeichen 05P12RFFTS) and by the \linebreak Helmholtz International
Center for FAIR (HIC for FAIR) within the framework of the LOEWE program
(Landesoffensive zur Entwicklung Wissenschaftlich-{\"O}konomischer
Exzellenz) launched by the State of Hesse. C.\ W.\ and A.\ M.\
acknowledge support by the Helmholtz Graduate School for Hadron and Ion
Research (HGS-HIRe), and the Helmholtz Research School for Quark Matter
Studies in Heavy Ion Collisions (HQM). Numerical computations have been
performed at the Center for Scientific Computing (CSC). H.\ v.\ H.\ has
been supported by the Deutsche Forschungsgemeinschaft (DFG) under grant
number GR 1536/8-1. C.\ W.\ has been supported by BMBF under grant
number 05P12RFFTS.

\begin{flushleft}
\bibliographystyle{epj}
\bibliography{qftbib}

\begin{thebibliography}{10}
\providecommand{\url}[1]{\texttt{#1}}
\providecommand{\urlprefix}{URL }
\providecommand{\eprint}[2][]{\url{#2}}

\bibitem{Friman:2011zz}
B.~Friman, C.~Hohne, J.~Knoll, S.~Leupold, J.~Randrup, et~al., Lect.Notes Phys.
  \textbf{814}, pp. 980 (2011),
  \urlprefix\url{http://dx.doi.org/10.1007/978-3-642-13293-3}

\bibitem{Kovacs:2006ym}
P.~Kovacs and Z.~Szep, Phys. Rev. D \textbf{75}, 025015 (2007),
  \urlprefix\url{http://dx.doi.org/10.1103/PhysRevD.75.025015}

\bibitem{Kovacs:2007sy}
P.~Kovacs and Z.~Szep, Phys. Rev. D \textbf{77}, 065016 (2008),
  \urlprefix\url{http://dx.doi.org/10.1103/PhysRevD.77.065016}

\bibitem{Gupta:2011ez}
U.~S. Gupta and V.~K. Tiwari, Phys. Rev. D \textbf{85}, 014010 (2012),
  \urlprefix\url{http://dx.doi.org/10.1103/PhysRevD.85.014010}

\bibitem{vanHees:2013qla}
H.~van Hees, C.~Wesp, A.~Meistrenko, and C.~Greiner, Acta Phys. Pol. Supp.
  \textbf{7}, 59 (2013),
  \urlprefix\url{http://dx.doi.org/10.5506/APhysPolBSupp.7.59}

\bibitem{Wesp:2014xpa}
C.~Wesp, H.~van Hees, A.~Meistrenko, and C.~Greiner, Phys. Rev. E \textbf{91},
  043302 (2015), \urlprefix\url{http://dx.doi.org/10.1103/PhysRevE.91.043302}

\bibitem{Greiner:2015tra}
C.~Greiner, C.~Wesp, H.~van Hees, and A.~Meistrenko, J. Phys. Conf. Ser.
  \textbf{636}, 012007 (2015),
  \urlprefix\url{http://dx.doi.org/10.1088/1742-6596/636/1/012007}

\bibitem{Kovacs:2016juc}
P.~Kov{\'a}cs, Z.~Sz{\'e}p, and G.~Wolf, Phys. Rev. D \textbf{93}, 114014
  (2016), \urlprefix\url{http://dx.doi.org/10.1103/PhysRevD.93.114014}

\bibitem{Stephanov:1999zu}
M.~A. Stephanov, K.~Rajagopal, and E.~V. Shuryak, Phys. Rev. D \textbf{60},
  114028 (1999), \urlprefix\url{http://dx.doi.org/10.1103/PhysRevD.60.114028}

\bibitem{Schaefer:2007pw}
B.-J. Sch{\"a}fer, J.~M. Pawlowski, and J.~Wambach, Phys. Rev. D \textbf{76},
  074023 (2007), \urlprefix\url{http://dx.doi.org/10.1103/PhysRevD.76.074023}

\bibitem{Skokov:2010uh}
V.~Skokov, B.~Friman, and K.~Redlich, Phys. Rev. C \textbf{83}, 054904 (2011),
  \urlprefix\url{http://dx.doi.org/10.1103/PhysRevC.83.054904}

\bibitem{Skokov:2010wb}
V.~Skokov, B.~Stokic, B.~Friman, and K.~Redlich, Phys. Rev. C \textbf{82},
  015206 (2010), \urlprefix\url{http://dx.doi.org/10.1103/PhysRevC.82.015206}

\bibitem{BraunMunzinger:2011ta}
P.~Braun-Munzinger, B.~Friman, F.~Karsch, K.~Redlich, and V.~Skokov, Nucl.
  Phys. A \textbf{880}, 48 (2012),
  \urlprefix\url{http://dx.doi.org/10.1016/j.nuclphysa.2012.02.010}

\bibitem{Schaefer:2011pn}
B.-J. Schaefer, Phys. Atom. Nucl. \textbf{75}, 741 (2012),
  \urlprefix\url{http://dx.doi.org/10.1134/S1063778812060270}

\bibitem{Morita:2012kt}
K.~Morita, V.~Skokov, B.~Friman, and K.~Redlich, Eur. Phys. J. C \textbf{74},
  2706 (2014), \urlprefix\url{http://dx.doi.org/10.1140/epjc/s10052-013-2706-1}

\bibitem{Nahrgang:2011mg}
M.~Nahrgang, S.~Leupold, C.~Herold, and M.~Bleicher, Phys. Rev. C \textbf{84},
  024912 (2011), \urlprefix\url{http://dx.doi.org/10.1103/PhysRevC.84.024912}

\bibitem{Nahrgang:2011mv}
M.~Nahrgang, S.~Leupold, and M.~Bleicher, Phys. Lett. B \textbf{711}, 109
  (2012), \urlprefix\url{http://dx.doi.org/10.1016/j.physletb.2012.03.059}

\bibitem{Herold:2013bi}
C.~Herold, M.~Nahrgang, I.~Mishustin, and M.~Bleicher, Phys. Rev. C
  \textbf{87}, 014907 (2013),
  \urlprefix\url{http://dx.doi.org/10.1103/PhysRevC.87.014907}

\bibitem{engquist1977absorbing}
B.~Engquist and A.~Majda, Proceedings of the National Academy of Sciences
  \textbf{74}, 1765 (1977)

\bibitem{Givoli2004319}
D.~Givoli, Wave Motion \textbf{39}, 319 (2004), new computational methods for
  wave propagation,
  \urlprefix\url{http://dx.doi.org/10.1016/j.wavemoti.2003.12.004}

\bibitem{PhysRevE.88.053308}
R.~M. Feshchenko and A.~V. Popov, Phys. Rev. E \textbf{88}, 053308 (2013),
  \urlprefix\url{http://dx.doi.org/10.1103/PhysRevE.88.053308}

\bibitem{Turk2010}
M.~J. Turk, B.~D. Smith, J.~S. Oishi, S.~Skory, S.~W. Skillman, T.~Abel, and
  M.~L. Norman, {ApJS} \textbf{192}, 9 (2010),
  \urlprefix\url{http://dx.doi.org/10.1088/0067-0049/192/1/9}

\bibitem{Abada:1994mf}
A.~Abada and J.~Aichelin, Phys. Rev. Lett. \textbf{74}, 3130 (1995),
  \urlprefix\url{http://dx.doi.org/10.1103/PhysRevLett.74.3130}

\bibitem{Greiner:1996md}
C.~Greiner and D.-H. Rischke, Phys. Rev. C \textbf{54}, 1360 (1996),
  \urlprefix\url{http://dx.doi.org/10.1103/PhysRevC.54.1360}

\end{thebibliography}
\end{flushleft}

\end{document}